\documentclass[aps,pra,twocolumn,showpacs,amsmath,amssymb,floatfix,footinbib,superscriptaddress]{revtex4-1}
\usepackage{graphicx}
\usepackage{amsmath}
\usepackage{latexsym}
\usepackage{amsmath}
\usepackage{amssymb}
\usepackage{float}
\usepackage{epstopdf}
\usepackage{bm}
\usepackage{amsfonts}
\usepackage[ansinew]{inputenc}
\usepackage[usenames,dvipsnames]{pstricks}
\usepackage{subfigure}
\usepackage{epsfig}
\usepackage{longtable}
\usepackage{url}
\usepackage{hyperref}
\usepackage{color}
\definecolor{darkblue}{rgb}{0.0,0.0,0.3}
\hypersetup{colorlinks,breaklinks,linkcolor=blue,urlcolor=blue,anchorcolor=blue,citecolor=blue}

\begin{document}
\baselineskip=14pt
\title{Excited-State Density-Functional Theory Revisited: on the Uniqueness, Existence, and Construction
of the Density-to-Potential Mapping}
\author{Prasanjit Samal}
\email{E-mail: psamal@niser.ac.in}
\affiliation{School of Physical Sciences, National Institute of Science Education \& Research,
Bhubaneswar 751005, INDIA.}
\author{Subrata Jana}
\affiliation{School of Physical Sciences, National Institute of Science Education \& Research,
Bhubaneswar 751005, INDIA.}
\author{Sourabh S. Chauhan}
\affiliation{School of Physical Sciences, National Institute of Science Education \& Research,
Bhubaneswar 751005, INDIA.}

\date{\today}
\begin{abstract}
The generalized constrained search formalism is used to address the issues concerning density-to-potential 
mapping for excited states in time-independent density-functional theory. The multiplicity of potentials 
for any given density and the uniqueness in density-to-potential mapping are explained within the framework 
of unified constrained search formalism for excited-states due to G\"{o}rling, Levy-Nagy, Samal-Harbola and 
Ayers-Levy. The extensions of Samal-Harbola criteria and it's link to the generalized constrained search 
formalism are revealed in the context of existence and unique construction of the density-to-potential 
mapping. The close connections between the proposed criteria and the generalized adiabatic connection
are further elaborated so as to keep the desired mapping intact at the strictly correlated regime. 
Exemplification of the unified constrained search formalism is done through model systems in order to 
demonstrate that the seemingly contradictory results reported so far are neither the true confirmation 
of lack of Hohenberg-Kohn theorem nor valid representation of violation of Gunnarsson-Lundqvist theorem 
for excited states. Hence the misleading interpretation of subtle differences between the ground and 
excited state density functional formalism are exemplified.
\end{abstract}
\maketitle

\section{introduction}
Since its advent, density-functional theory (DFT) \cite{hk,ks,cs1,cs2,cs3,cs4,cs5,py,dg,dj} is routinely 
applied for calculating the electronic, magnetic, spectroscopic and thermodynamic properties of atoms, 
molecules and materials in ground and excited states. In the last couple of decades, studying excited-
states employing DFT has become the main research interest \cite{py,dg,dj,gl1,gl2,zr,vb,at,at2,at3,zls,
wh,pl,gok,mlp,anp,gor1,gor2,gor3,ln1,ln2,ln3,pgg,su,fug,anqc,rvl,cjcs,pplb,lps,av,jked,gj,thesis,rvl,cc,
rg,dg,kd,sd,gk,cau,est}. Thus one of the most natural approach to do excited-state DFT is to adopt the 
time-independent density functional formalism \cite{anp,anqc,an,thesis} in which the individual excited-state 
energies are determined from the stationary states of the energy density functional. However, the question 
is whether there exists any such functional(s) for excited states analogous to the ground-state. Not only 
energy functionals but  also the most fundamental and essential requirement for excited-state density
functional theory ($e$DFT) is to establish the one-to-one mapping similar to the Hohenberg-Kohn theorem 
which is the main intent of the present work. Although the issue of density $\rho(\vec r)$ to potential 
$\hat v(\vec r)$ mapping for excited states has been addressed in the past \cite{sahni,sahni1,sahni2,gb,
har1,lpls}, but the question still remains unanswered. So the current work will answer the critiques of
density-to-potential mapping based on the generalized/unified constrained search(CS) due to Perdew-Levy(PL) 
\cite{pl}, G\"{o}rling \cite{gor1,gor2,gor3}, Levy-Nagy(LN) \cite{ln1,ln2,ln3}, Samal-Harbola(SH) 
\cite{shcpl,shjp1,shjp2,shjp3} and Ayers-Levy-Nagy \cite{al,aln,pair}. 

In the present work, we will critically analyse and make furtherance to the $e$DFT ideas proposed 
by Samal and Harbola \cite{shjp2,thesis}. According to it, (i) the CS approach can be extended 
to excited-state in the light of the stationary state formalism of G\"{o}rling \cite{gor1,gor2,gor3} 
and variational $e$DFT formalism by Levy-Nagy \cite{ln1,ln2,ln3}; (ii) within the variational $e$DFT 
formalism, the construction of the Kohn-Sham(KS) system by comparing only the ground-state density 
is insufficient and can't explain the existence of multiple potentials; (iii) the density-to-potential 
mapping in $e$DFT can be achieved through the following criteria: compare the ground states of the true 
and KS system energetically such that it can account for the most close resemblance of the densities in 
a least square sense. SH showed it by comparing the expectation value of the original ground-state 
KS Hamiltonian (obtained using the Harbola-Sahni \cite{hs} exact exchange potential) with that of 
the alternative KS systems. Finally, the kinetic energy of true and KS system need to be kept closest. 
This is also another way of comparing the ground states based on the differential virial theorem(DVT) 
\cite{hm}; (iv) the CS approach is capable of generating all the potentials for a given excited state 
density and at the same time uniquely establishes the density-to-potential mapping.

The work is organized as follows. In Sec.II, the generalized/unified CS $e$DFT will be briefly discussed 
from the prospective of density-to-potential mapping. It will be shown that there exist multitude of 
potentials for a given density. In Sec.III, furtherance of SH $e$DFT will be presented in order to show 
the existence and unique construction of the desired density-to-potential mapping. In this, we will show, 
how the proposed $e$DFT is also consistent with the generalized adiabatic connection(GAC) KS formalism 
\cite{gl1,gor1,gor2,gor3,gacj,gacl,gac1,gac2,gac3,gac4,gac5,gac6,gac7} and in principle applicable to 
(non-)coulombic densities. In Sec.IV, we will show the existence of multiple potentials for given ground 
or lowest excited states can never be ruled out even within Li. et al.\cite{lpls} demonstration of 
Gunnarsson and Lundqvist(GL) theorem \cite{gl1,gl2}. However, based on the theories presented in Sec.II 
\& III, these seemingly contradictory results will be explained in order to justify the non-violation of
Hohenberg-Kohn(HK) \cite{hk} and GL theorems. Thus the density-to-potential mapping will be demonstrated 
within \cite{lpls} approach by making use of the unified $e$DFT for the two model systems (i.e. $1D$ 
quantum harmonic oscillator(QHO) with finite boundary and infinite well external potentials). For 
completeness, in Sec. V, same set of model systems will be used to exemplify density-to-potential mapping 
based on the CS formalism \cite{zp}. Finally, we will provide firm footing to density-to-potential mapping 
based on the proposed criteria of $e$DFT.

\section{Unified Constrained-Search Formulation of \lowercase{e}DFT}
Although in principle the ground-state CS formalism \cite{cs1,cs2,cs3,cs4,cs5} has all the information 
about the excited-states, the desired density-to-potential mapping for individual excited-states are 
not so trivial and straightforward. To do so, series of attempts being made based on the original CS 
approach \cite{ln1,ln2,ln3,pl,gor1,gor2,gor3,al,anp,aln,mlp,shcpl,shjp1,shjp2,shjp3,est,pair}. In the 
recent past, the form of functional for ground state (both for degenerate and non-degenerate) has been 
extended \cite{shjp1,shjp2,shamim,hem1,hem2,hem3} to study the excited states. Now we will briefly 
describe how the generalized CS formalism explains the existence of multiple potentials for any given 
fermionic density without hindering the density-to-potential mapping. 
 
Let's consider $N$ fermions trapped in a local external potential $\hat v_{\text{ext}}(\vec r)$, 
described by the Hamiltonian  
\begin{equation}
\hat H[\hat v;N] = \hat T + \hat V_{\text{ee}} + \sum_{i=1}^N\hat v_{\text{ext}}(\vec r_i),
\label{gcs1}
\end{equation}
where $\hat T$ and $\hat V_{ee}$ are the kinetic and electron-electron interaction operators with the 
corresponding stationary states are given by
\begin{equation}
\hat H[\hat v(\vec r), N] \Psi_k(\vec r) = E_k[\hat v(\vec r), N] \Psi_k(\vec r)~,
\label{gcs2}
\end{equation}
where $\hat v_{\text{ext}}(\vec r) \equiv \hat v(\vec r)$. In Eq.(\ref{gcs2}), $\Psi_k(\vec r) \equiv 
\Psi_k[\hat v(\vec r), N] $ are the pure state $v-$representable stationary quantum states i.e. it is 
coming from the solution of the Schr\"odinger equation. But for $N-$representable densities (i.e. $\int 
\rho(\vec r) d\vec r = N$) and therefore wavefunctions (i.e. $\int {\Psi[N]}^2 d\vec r = N$), similar 
to the HK universal functional there exists an analogous functional which is stationary w.r.t all the 
variations that do not change the density (i.e. $\delta_{\Psi \to \rho}$) and is given by
\begin{equation}
Q^S[\rho;N] = \delta_{\Psi[N] \to \rho(\vec r)}\langle\Psi | \hat T + \hat V_{\text{ee}} | \Psi\rangle~.
\label{gcs3}
\end{equation}
Now according to the Perdew and Levy extremum principle \cite{pl} and generalized CS formalism \cite{gor2,
shjp2,al}, the energy of the $k^{th}$ excited state is given by
\begin{equation}
E_k = E[\rho_k;N] = Q^S[\rho_k;N] + \int\rho_k(\vec r) {v}_{\text{ext}}(\vec r) d\vec r .
\label{gcs4}
\end{equation} 
In Eq.(\ref{gcs4}), the minimization occurs only over G\"{o}rling's stationary-state functional $Q^S[\rho_k]$ 
and the corresponding wavefunctions are given by
\begin{equation}
\Psi_k^S = \Psi^S[\rho_k,N] = \arg\min_{\Psi[N] \to \rho_k}\langle\Psi[N]|\hat T + \hat V_{\text{ee}}|
\Psi[N]\rangle .
\label{swfc}
\end{equation} 
On the other hand, in the LN \cite{ln1,ln2,ln3} variational constrained minimization approach for 
excited-states leads to the $k^{th}$ stationary state energy 
\begin{eqnarray}
E_k[\rho,\rho_0] &=& \min_{\rho[\hat v] \to N}\Big\{\int \rho(\vec r) {v}_{\text{ext}}(\vec r) 
d\vec r + F[\rho,\rho_0]\Big\} \nonumber\\
&=& \int \rho_k(\vec r) {v}_{\text{ext}}(\vec r) d\vec r + F[\rho_k,\rho_0]~,
\label{gcs5}
\end{eqnarray} 
where $\rho_0$ is the ground state density of the system under consideration. The LN energy density 
functional differs from the HKS ground-state and the stationary state $e$DFT functional due to the 
bifunctional $F[\rho,\rho_0]$, which is defined by
\begin{eqnarray}
F_k[\rho,\rho_0] &=& \min_{\Psi[N] \to \rho, {\langle\Psi[N]|\Psi_j[\hat v;N]
\rangle=0, j < k}} \langle\Psi|\hat T + \hat V_{\text{ee}}|\Psi\rangle \nonumber \\
&=& F[\rho_k,\rho_0] ,
\label{gcs6}
\end{eqnarray}
for the $k^{th}$ excited state. So the energy of the $k^{th}$ excited state can be re-expressed as
\begin{eqnarray}
E_k[\rho,\rho_0] = \min_{\rho[\hat v] \to N} \Big\{\int \rho_k(\vec r) { v}_{\text{ext}}(\vec r) d\vec r 
+ \nonumber\\ \min_{\Psi[N] \to \rho, {\langle\Psi[N]|\Psi_j[\hat v;N] \rangle=0, j < k}} \langle\Psi|
\hat T + \hat V_{\text{ee}}|\Psi\rangle\Big\}~
\label{gcs7}
\end{eqnarray}
with the minimizing wavefunction denoted by
\begin{equation}
\Psi_k^{LN}[\hat v;N] = \arg \min_{\Psi[N] \to \rho, {\langle\Psi[N]|\Psi_j[\hat v;N]
\rangle=0, j < k}} \langle\Psi|\hat T + \hat V_{\text{ee}}|\Psi\rangle .
\label{vwfc}
\end{equation}
In the LN bifunctional $F[\rho,\rho_0]$, if $\rho = \rho_0$ then the functional reduces to HK universal
functional and the same holds true for $Q^S[\rho]$. Also $F[\rho,\rho_0]$ is the generalization of the $e$DFT 
stationary state functional $Q^S[\rho]$ as described in the $\it Theorems 4,~5~\&~6$ of \cite{al}. These 
theorems are in fact an artifact of the orthogonality constraint. Since all the lower states $\Psi_j[\hat 
v;N] (j < k)$ are determined from the external potential $\hat v_{\text{ext}}$ (which is a unique functional 
of ground state density $\rho_0$ according to HK \cite{hk} theorem), implies that the ground state density 
plays an important role in LN-formalism. So, in principle one can also write the excited-state density 
bifunctional as $F_k[\rho,{\hat v}_{\text{ext}}]$ instead of $F_k[\rho,\rho_0]$. If the electronic densities 
are $v-$representable then Eq.(\ref{gcs6}) modifies to 
\begin{equation}
E_k[\hat v_{\text{ext}};N] = \int \rho_k(\vec r) {v}_{\text{ext}}(\vec r) d\vec r + 
F_k[\rho,{\hat v}_{\text{ext}}].
\label{gcs8}
\end{equation}
From the generalized CS energy functionals for any eigendensity given by the Eq.(\ref{gcs4}) \& Eq.(\ref{gcs8})
there exist multiple potential functions \cite{shjp2,thesis,al}. In general, these generalized multiple local 
external potentials can be obtained through the Euler Lagrange equation   
\begin{eqnarray}
\frac{\delta}{\delta\rho}\Big[E_k - \mu\Big\{\int \rho_k(\vec r) d\vec r - N \Big\}\Big] = 0 \\
 v_{\text{ext}}(\vec r) = \mu - \Big(\frac{\delta F[\rho,\rho_0]}{\delta\rho}\Big){\Big |}_{\rho = \rho_k} \\
{\text OR}~~ 
w_{\text{ext}}(\vec r) = \mu - \Big(\frac{\delta Q^S[\rho(\vec r)]}{\delta\rho}\Big){\Big |}_{\rho = \rho_k},
\label{gcs9}
\end{eqnarray}
where $\mu = \left(\frac{\delta E[\rho_e]}{\delta\rho} \right)_N$. The actual potential is one of these which
should be uniquely mapped to the given density as will be shown in the following sections. In particular, the 
local external potentials will be identical $v_{\text{ext}}(\vec r) = w_{\text{ext}}(\vec r)$ iff the density 
$\rho_k(\vec r)$ is pure state $v$-representable and $E_k$ is the corresponding eigen energy. Now to obtain  
the KS like equation for the generation of $\rho_k$ and  to obtain $E_k$, one needs to first construct a non
-interacting system with some external potential ${\hat v}'_{\text{ext}}$ such that it's $m^{th}$ excited 
state density $\rho^{{\hat v}'_{\text{ext}}}_m(\vec r)$ (say) may be the same as $\rho_k(\vec r)$ of the 
original system ${\hat v}_{\text{ext}}$. In stationary-state $e$DFT \cite{gor2,shjp2,thesis}, this is done 
by generalized adiabatic connection (GAC) \cite{gl1,gor1,gor3,gacj,gacl,gac1,gac2,gac3,gac4,gac5,gac6,gac7}. 
Whereas, in LN variational $e$DFT \cite{ln1,ln2,ln3,al}, this is done by the constrained minimization of the 
expectation value $\langle\Psi[{\hat v}'_{\text{ext}},\rho^{{\hat v}'_{\text{ext}}}_m(\vec r)]|\hat T + 
{\{{\hat V}_{\text{ee}}=0\}}|\Psi[{\hat v}'_{\text{ext}},\rho^{{\hat v}'_{\text{ext}}}_m(\vec r)]\rangle$, 
where $\Psi[{\hat v}'_{\text{ext}}, \rho^{{\hat v}'_{\text{ext}}}_m(\vec r)]$ gives the desire density of 
interest. Out of many such non-interacting $\Psi[{\hat v}'_{\text{ext}},\rho^{{\hat v}'_{\text{ext}}}_m(
\vec r)]$~s (different systems), the unique one is chosen whose ground-state density $\rho^{{\hat v}'_{
\text{ext}}}_0(\vec r)$(say) resembles with the ground-state density $\rho^{{\hat v}_{\text{ext}}}_0(\vec r)$ 
of the original system ``most closely in a least-square sense"(i.e. the LN criterion). The matching of the 
ground-state densities actually matches the external potentials ${\hat v}'_{\text{ext}}$ and ${\hat v}_{
\text{ext}}$ according to the HK theorem \cite{hk}. But the difference occurs between the kinetic energies 
of the two systems. As matter of which, the discrepancy in the $\rho \Longleftrightarrow \hat v$ mapping 
arises because the LN criterion strictly depends upon the behavior of the bifunctional.

\section{Proposed Constrained-Search Formulation of \lowercase{e}DFT}
The CS formulation described in the previous section implies that the content of the excited state 
functionals $Q^S[\rho_e]$ and $F[\rho_e,\rho_0]$ differs from the HK universal functional $F[\rho]$ 
except their stationarity with respect to variation in the external potential. Actually, only in 
the case of ground-state, all the three functionals are identical to one another and in general 
there exists a close link between G\"{o}rling $Q^S[\rho_e]$ and Levy-Nagy $F[\rho_e,\rho_0]$ \cite{al}. 
So in the unified $e$DFT formalism, for a given excited-state eigendensity $\rho_e(\vec r)$, both 
$Q^S[\rho_e]$ and $F[\rho,\hat v_{ext}]$ are stationary about the corresponding $\hat v_{ext}$ which 
also holds for the desired excited-state $\Psi_k^S \equiv \Psi_k^{LN}$ \cite{shjp2,thesis,al}. 
Now due to the presence of orthogonality constraint in $F[\rho,\hat v_{ext}]$, several choices for 
the set of low lying states can be made to which $\Psi_k^{LN}$ will be orthogonal and for each choice, 
there may exists a generalized potential function $\hat w_{ext}$. So some extra deciding factors are 
required for setting up the $\rho \Longleftrightarrow \hat v$ mapping which is the intent of the current 
section. 

Now resorting back to the work of Samal-Harbola \cite{shjp2}, we would also like to re-emphasis that 
the direct or indirect comparison of ground states are not sufficient to establish the $\rho(\vec r) 
\Longleftrightarrow {\hat v}_{ext}(\vec r)$ mapping or to construct the KS system for excited-states 
\cite{shcpl}. Given the discussions on unified CS $e$DFT in the previous section, we now present a 
consistent approach to address the density-to-potential mapping issues. Fundamentally rigorous and 
crucial tenets of the proposed \lowercase{e}DFT are: ({\it i}) There exist ways for mapping an 
excited-state density $\rho_e(\vec r)$ to the corresponding many-electron wavefunction $\Psi(\vec r)$ 
which in turn maps to the external potential $\hat v_{ext}(\vec r)$ through the $\rho$-stationary 
wavefunctions \cite{gor2,shjp2,al}. In this, the wavefunction depends upon the ground-state density 
$\rho_0$ implicitly. ({\it ii}) The KS system is to be defined through a comparison of the kinetic 
energy, ground-state density and variation of the energy w.r.t. symmetry of the excited-states.

The claim is, unified CS approach can provide the mapping from an excited-state density $\rho_e(\vec r)$ 
to many-body wavefunction. Stationary state formalism \cite{gor2,shjp2} provides a straightforward method 
of mapping $\rho_e(\vec r) \Longleftrightarrow {\hat v}_{ext}(\vec r)$, just by making sure whether $\langle
\Psi_k|\hat T + \hat V_{\text{ee}}|\Psi_k\rangle$ is stationary or not, subject to the condition that 
$\Psi_k$ gives $\rho_e$. But \cite{gor2,shjp2,thesis,al} shows that different $\Psi_k(\vec r)$s correspond 
to potentials ${\hat v}^k_{ext}(\vec r)$. The same problem also pervades through the variational $e$DFT 
approach as proposed by LN \cite{ln1,ln2,shjp2}. Thus unified CS gives, many different wavefunctions 
$\Psi_k(\vec r)$ and the corresponding external potential ${\hat v}^k_{ext}(\vec r)$ can be associated 
with a given density. Now if in addition to the excited-state density we also have the ground-state 
information $\rho_0$, then ${\hat v}_{ext}(\vec r)$ can be uniquely determined out of all possible multiple
potentials ${\hat v}^k_{ext}(\vec r)$. Hence with the knowledge of $\rho_0$, it is quite trivial to select 
a particular $\Psi$ that belongs to a given $\left[\rho_e,\rho_0\right]$ combination by comparing 
${\hat v}^k_{ext}(\vec r)$ with the actual ${\hat v}_{ext}(\vec r)$. Alternatively, one can think of it as 
finding $\Psi$ variationally for a $\left[\rho_e,{\hat v}_{ext}\right]$ combination. Its because the knowledge 
of $\rho_0$ and ${\hat v}_{ext}$ is equivalent. Now with the above information, the bifunctional $F[\rho_e,
\rho_0]$ can be redefined as
\begin{equation}
F[\rho_e,\rho_0] \; = \;
\langle \Psi[\rho_e,\rho_{0}] | \hat{T} \;+\; \hat{V}_{ee} | \Psi[\rho_e,\rho_{0}]\rangle .
\label{cdpm1}
\end{equation}
The above theoretical formulation is similar to that of LN \cite{ln1} but avoids the orthogonality 
constraint imposed by LN formalism. This is because, the densities for different excited states for 
a given ground-state density $\rho_0$ (that corresponds to a unique external potential ${\hat v}_{ext}$) 
can be found in following manner: take a density and search for $\Psi$ that makes $\langle\Psi|
\hat T + \hat V_{ee}|\Psi\rangle$ stationary and simultaneously make sure whether the corresponding 
potential ${\hat w}_{ext}$ $\Big(i.e.~ w_{ext} = -~\frac{\delta F[\rho,\rho_0]}{\delta\rho}{\Big|}_{\rho 
= \rho_e} \Big)$ resembles the given $\rho_0$ ( or ${\hat v}_{ext}$); if not, search for another 
density and repeat the procedure until the correct $\rho$ is found. Thus it is clear that excited 
state orbitals $\Psi$ are now functional of $[\rho_e,\rho_0]$. So the correct density $\rho$ is 
excited state density of the potential and the $\Psi$ obtained in this method is also excited state 
wavefunction corresponding to that potential and density. After finding the correct density $\rho_e$, 
make a variation over it so that $(\rho_e \to \rho_e + \delta\rho)$ and again perform the CS to find 
$\Psi[\rho_e + \delta\rho;\rho_0]$. In this case, choose that $({\hat w}_{\text{ext}} + \delta 
{\hat w}_{\text{ext}})$ which converges to ${\hat v}_{\text{ext}}$ as $\delta\rho \to 0$. 

The above propositions for the excited-states in terms of their densities are quite reasonable, 
particularly because it's development is parallel to that for the ground-state DFT. On the other 
hand, to construct a Kohn-Sham \cite{ks} system for a given density is not so trivial; and to 
carry out accurate calculations for excited-states, it is of prime importance to construct a KS 
system. Further, a KS system will be meaningful if the orbitals involve in an excitation match 
with the corresponding excitations in the true system. Samal-Harbola \cite{shjp2} have shown that 
the KS system constructed using the Levy-Nagy criterion fails in this regard. But using the form 
of the functional above a KS system can be defined for excited state. Actually, the state dependence 
of the excited-state exchange-correlation functional \cite{shjp1,shamim,hem1,hem2,hem3} leads to 
the discrepancies while one compares the ground-states either direct or indirect manner. But in 
principle, obtaining a KS system is plausible. Now by defining the non-interacting kinetic energy 
$T_s\left[\rho_e,\rho_0\right]$ and using it to further define the exchange-correlation functional as 
\begin{equation}
E_{xc}[\rho_e,\rho_0] = F[\rho_e,\rho_0] - E_{\text{H}}[\rho_e] - T_s[\rho_e,\rho_0],
\label{cdpm4}
\end{equation}
solves the purpose. So the Euler equation for the excited-state densities becomes
\begin{equation}
{v}_{ext} = \mu - \Big\{\frac{\delta T_s\left[\rho_e,\rho_0\right]}{\delta\rho(\vec r)} + {V}_{\text
{H}}[\rho_e] + \frac{\delta E_{xc}\left[\rho_e,\rho_0\right]}{\delta\rho(\vec r)}\Big\}~,
\label{cdpm5}
\end{equation}
which is equivalent to solving the KS equation
\begin{equation}
\left\{- \frac{1}{2}\nabla^2 + {\hat v}_{s}(\vec r) \right\}\Psi_i(\vec r) = \varepsilon_i \Psi_i(\vec r)~,
\label{cdpm6}
\end{equation}
where
\begin{equation}
v_{s}(\vec r) = v_{ext}(\vec r) + \frac{\delta \Big\{F\left[\rho,\rho_0\right] - T_{s}\left[\rho,\rho_0
\right]\Big\}}{\delta\rho(\vec r)}{\Big|}_{\rho(\vec r) = \rho_e[v_{ext}(\vec r)]} ~.
\label{cdpm7}
\end{equation}
In ground state DFT, one can easily find the $T_s[\rho_0]$ by minimizing the kinetic energy for a given 
density; here $T_s[\rho_0]$ for a given density is obtained by occupying the lowest energy orbitals for 
a non-interacting system. But in $e$DFT, to define $T_s\left[\rho_e,\rho_0\right]$ is not easy, as for 
the excited-states it is not clear which orbitals to occupy for a given density. Particularly because 
a density can be generated by many different configurations of the non-interacting systems. Levy-Nagy 
select one of these systems by comparing the ground-state density corresponding to the excited-state 
non-interacting system with the true ground-state density. However, LN criterion is not satisfactory 
as pointed out by Samal and Harbola \cite{shcpl}. The reason of this discrepancy is due to the inconsistency 
of the ground-state density of an excited state KS system with the true ground-state density. The ground
-state density corresponding to the excited-state KS system is not same as the ground-state density of 
the true system. This means the desired state is not associated with ${\hat v}_{\text{ext}}(\vec r)$, 
rather it comes from a different local potential ${\hat v}'_{\text{ext}}(\vec r)$. To settle this 
inconsistency, KS system must be so chosen that it is energetically very close to the original system 
and it can be ensured through the following criterion. {\it Criterion I: the non-interacting kinetic 
energy} $T_s[\rho_e,\rho_0]$ {\it obtained through the CS need to be very close to the actual} 
$T[\rho_e,\rho_0]$, where $T_s[\rho_e,\rho_0]$ and $T[\rho_e,\rho_0]$ are defined as 
\begin{eqnarray}
T_s[\rho_e,\rho_0]&=&\min_{\Phi \to \rho_e} \langle\Phi|\hat T + \underbrace{\hat V_{\text{ee}}=0}|
\Phi\rangle\nonumber\\
T[\rho_e,\rho_0]&=&\min_{\Psi \to \rho_e}\langle\Psi|\hat T + \hat V_{\text{ee}} |\Psi\rangle .
\label{cdpm8} 
\end{eqnarray}
So defining $\Delta T = T - T_s$ smallest not only ensures that DFT exchange-correlation energy 
remains closer to the conventional quantum mechanical exchange-correlation energy but also keeps the 
structure of the KS potential appropriate for the desired excited-state which is shown below. Based 
on the DVT \cite{hm}, it can be argued how for a given density $\rho_e$ one can have different exchange
-correlation $\hat v_{xc}$ and external ${\hat v}_{\text{ext}}$ potentials. According to DVT, the exact 
expression for the gradient of the external potential (for interacting system) for a given excited-state 
density $\rho_e$ is 
\begin{eqnarray}
-\nabla {\hat v}_{\text{ext}} &=& -\frac{1}{4\rho_e(\vec r)}\nabla\nabla^2\rho_e(\vec r) + 
\frac{1}{\rho_e(\vec r)}\vec Z(\vec r;\varGamma_1(\vec r;\vec r'))\nonumber\\
& + &\frac{2}{\rho_e(\vec r)}\int[\nabla {\hat u}(\vec r,\vec r')]\varGamma_2(\vec r,\vec r')d\vec r'~,
\label{dvt1}
\label{cdpm9}
\end{eqnarray}
where ${\hat u} = \frac{1}{|\vec r - \vec r'|}$. This equation represents an exact relation between 
the gradient of the external potential ${\hat v}_{\text{ext}}$, the $e-e$ interaction potential 
${\hat u}(\vec r,\vec r')$ and the density matrices $\rho(\vec r)$, $\varGamma_1(\vec r;\vec r')$ 
and $\varGamma_2(\vec r,\vec r')$. The vector field $\vec Z$ in Eq.(\ref{cdpm9})is related to the 
kinetic-energy density tensor via 
\begin{equation}
 Z_\alpha[\vec r;\varGamma_1(\vec r;\vec r')] = \Big[\frac{1}{4}\Big(\frac{\partial^2}
{\partial r'_\alpha\partial r''_\beta} + \frac{\partial^2}{\partial r'_\beta\partial r''_\alpha} 
\Big)\varGamma_1(\vec r';\vec r'')\Big]_{\vec r'=\vec r''=\vec r}           
\label{cdpm10}
\end{equation}
So, $\vec Z$ can be called a ''local'' functional of $\varGamma_1$. Similarly, for KS potential 
Eq.(\ref{cdpm9}) reduces to
\begin{equation}
\nabla {\hat v}_{\text{KS}} = -\frac{1}{4\rho_e(\vec r)}\nabla\nabla^2\rho_e(\vec r) + \frac{1}
{\rho_e(\vec r)}\vec Z_{\text{KS}}(\vec r;\varGamma_1(\vec r;\vec r')) . 
\label{cdpm11}
\end{equation}
As a given ground-state density $\rho_0$ fixes the external potential uniquely via HK theorem, 
which implies that $\rho$, $\varGamma_1$ and $\varGamma_2$ are also fixed from Eq.(\ref{cdpm9}).
Since the density matrices generated by some eigenfunction $\Psi$ of the Hamiltonian $\hat H$. So 
the fixed pair of excited-state and ground-state density i.e. $[\rho_e,\rho_0]$ may be arising 
from different configurations $-$ different configurations can be thought of as arising from 
different external potential or different exchange-correlation potential and this is due to the 
different $\varGamma_1$ and $\varGamma_2$ for a fixed $\rho_e$. Suppose a given density $\rho_e$ 
is generated through an $i^{th}$ KS system, then
\begin{equation}
\nabla {\hat v}_{\text{KS}}^i = -\frac{1}{4\rho_e(\vec r)}\nabla\nabla^2\rho_e(\vec r) + \frac{1}
{\rho_e(\vec r)}\vec Z^i_{\text{KS}}(\vec r;\varGamma^i_{1 ({\text{KS}})}(\vec r;\vec r'))~.
\label{cdpm12}
\end{equation}
If the density is generated through a $j^{th}$ external potential then
\begin{eqnarray}
-\nabla {\hat v}_{\text{ext}}^j &=& -\frac{1}{4\rho_e(\vec r)}\nabla\nabla^2\rho(\vec r) + \frac{1}
{\rho(\vec r)}\vec Z^j(\vec r;\varGamma^j_1(\vec r;\vec r'))\nonumber\\
& + &\frac{2}{\rho_e(\vec r)}\int[\nabla u(\vec r,\vec r')]\varGamma^j_2(\vec r,\vec r')d\vec r'~.
\label{cdpm13}
\end{eqnarray}
As a matter of which
\begin{eqnarray}
- \nabla {\hat v}_{\text{xc}} = \frac{\vec Z_{\text{KS}}(\vec r;\varGamma_1(\vec r;\vec r')) - 
\vec Z(\vec r;\varGamma_1(\vec r;\vec r'))}{\rho_e(\vec r)} \nonumber + \\
\frac{\int[\nabla {\hat u}(\vec r,\vec r')][\rho_e(\vec r) \rho_e(\vec r') - \varGamma_2(\vec r,
\vec r')] d\vec r'}{\rho_e(\vec r)}
\label{cdpm-1}
\end{eqnarray}
becomes
\begin{equation}
-\nabla {\hat v}_{xc}^{ij} = \frac{\vec Z^i_{\text{KS}}-\vec Z^j}{\rho(\vec r)} + \vec\varepsilon_
{\text{xc}}^j~,
\label{cdpm14}
\end{equation}
where $\vec\varepsilon_{\text{xc}}^j$ is the field due to the Fermi-Coulomb hole of the $j^{th}$ 
system $[\varGamma_2^j]$ . So the kinetic energy difference between the true system and KS system is
given by
\begin{equation}
\Delta T = \frac{1}{2}\int\vec r.\Big\{\vec Z_{\text{KS}}\Big(\vec r;[\varGamma_{1({\text{KS}}}]
\Big) - \vec Z\Big(\vec r;[\varGamma_1]\Big)\Big\} d\vec r .
\label{cdpm15}
\end{equation}
This difference should be kept the smallest for the true KS system so that it gives the KS 
system consistent with the original system. As a matter of which, we conclude that one way 
to establish the $\rho_e \Longleftrightarrow {\hat v}_{\text{ext}}$ mapping via the LN 
formalism \cite{ln1,ln2,ln3} is: if among the several potentials $-$ which have the same excited
-state density, one can choose the correct KS potential by comparing the ground-state density 
i.e. keep that KS-potential whose ground-state density resembles with the true ground-state 
density. Keeping the ground-state density close we actually keep the external potential fixed 
via HK theorem. Thus LN criterion is exact for non-interacting system as there is no interaction, 
so the ground-state density match perfectly. 

This proposal of LN for $\rho_e \Longleftrightarrow {\hat v}_{\text{ext}}$ mapping was carried by Samal 
and Harbola \cite{shjp2} but they argued in a slightly different way. They proposed that both for 
interacting and non-interacting case among all the multiple potentials, choose the correct KS potential 
whose ground-state density differ from the exact ground-state ``most closely by least-square sense''
which is done in the following manner. If $\rho_0(\vec r)$ is the exact ground state density and 
$\tilde{\rho_0}(\vec r)$ is that of the KS system (OR the alternate potentials ${\hat w}_{\text{ext}}$) 
then SH proposition can be further improved intuitively. {\it Criterion II: the mean square distance 
between $\rho_0(\vec r)$ and $\tilde{\rho_0}(\vec r)$ should remain very close to zero}. Thus
\begin{equation}
\Delta[\rho_0(\vec r),\tilde{\rho_0}(\vec r)] = \min_{v[\tilde{\rho_0},\rho_e]} \left\{\int_
\infty\left|[\rho_0(\vec r) - \tilde{\rho_0}(\vec r)]\right|^2 d \vec r \right\}^\frac{1}{2} \geq 0~,
\label{cdpm16}
\end{equation}
where the integration is carried out in the Sobolev space. This criterion is more appropriate in the
context of $\rho_e \Longleftrightarrow {\hat v}_{\text{ext}}$ than the one proposed by \cite{shjp2,aln}.
The criterion as given in Eq.(\ref{cdpm16}) will be fully satisfied if one makes use of the excited
state functionals \cite{shjp1,thesis,shamim,hem2,hem3}. Otherwise it may fail in certain situations
as pointed out by SH \cite{shjp2}. 

Instead of sticking to the {\it Criterion I \& II}, one can even go beyond the same through {\it Criterion 
III: compare the ground states of the true and alternate systems energetically}. It can be done in the 
following manner in order to select the KS system for a given density. The alternative approach is to compare 
the ground-state expectation value of the KS system and the true system, instead of comparing their 
ground-state densities and kinetic energies. The procedure for comparing ground-state energy level is the
following. First solve the exact DFT equation (say Harbola-Sahni \cite{hs} etc) for ground-state of the 
true system and obtain the ground-state of KS Hamiltonian $H_0$. If the expectation value of the ground 
state Hamiltonian of the true system is $\langle H_0 \rangle_{\text{true}}$ and that of the KS system is 
$\langle H_0 \rangle_{\text{KS}}$, then one need to choose that KS system whose $\langle H_0 \rangle_
{\text{KS}}$ $\simeq$ $\langle H_0 \rangle_{\text{true}}$. These criteria are well connected to the GAC-KS
\cite{gl1,gor1,gor2,gor3,gacj,gacl,gac1,gac2,gac3,gac4,gac5,gac6,gac7} as discussed below.

Since GAC-KS in principle helps for the self-consistent treatment of excited states and could be considered 
as a plausible extension of HK theorem to the same. So now the furtherance of the propositions made by SH 
\cite{shjp2} as discussed previously will be justified within the GAC-KS.  Indeed, relying on the principles 
of GAC-KS, unified CS formalism along with the SH criteria can also establish the density-to-potential mapping 
at the strictly correlated regime which will be shown below. In GAC, the $\lambda$ dependent Hamiltonian 
which is also used in the PL extremum principle is given by
\begin{equation}
 \hat H_\lambda[\hat v, N] = \hat T + \lambda \hat V_{ee} + \sum_{i=1}^N \hat v(\vec r_i) ,
 \label{gaceq1}
\end{equation}
with the corresponding equation of state
\begin{equation}
 \hat H_\lambda[\hat v,N] \Psi_\lambda[\hat v,N] = E_\lambda[\hat v,N] \Psi_\lambda[\hat v,N] ,
 \label{gaceq2}
\end{equation}
where $\lambda$ is the coupling constant with $0 \leq \lambda \leq 1$ allowing the electron-electron 
interaction to be triggered. Unlike the adiabatic connection (AC)-DFT, the external potential
$\hat v(\vec r)$, is independent of $\lambda$. Analogous to the Levy-Lieb CS functionals, the GAC 
for the conjugate density functionals $F_\lambda[\rho]$ (density fixed AC) and $E_\lambda[\hat v]$ 
(potential fixed AC) are given by
\begin{equation}
F_{\lambda = 1}[\rho] = F_{\lambda = 0}[\rho] + \int_0^1 \frac{dF_\lambda[\rho]}{d\lambda}~d\lambda ~,
\label{gaceq3}
\end{equation}
\begin{equation}
E_{\lambda = 1}[\hat v] = E_{\lambda = 0}[\hat v] + \int_0^1 \frac{dE_\lambda[\hat v]}{d\lambda}~d\lambda ~.
\label{gaceq4}
\end{equation}
Similar to Eq.(\ref{gaceq3}) and (\ref{gaceq4}), the excited-state functionals $T_\lambda[\rho,\rho_0]$, 
$Q^S_\lambda[\rho]$, $F_\lambda[\rho,\rho_0]$ and $E_\lambda[\rho,\rho_0]$ can be defined. Upon finding 
these $e$DFT functionals, one can define the GAC by starting at a $\rho$ stationary wavefunction for 
$\lambda = 1$ and then by gradually turning off ($\lambda = 0$) the electron-electron interaction. Thus 
the $\rho$-stationary wavefunctions for $0 \leq \lambda \leq 1$ will form the GAC in $e$DFT. Since the 
$\rho$-stationary wave functions for a given $\rho$ are numerable and the adiabatic connections do not 
overlap with each other, states $\Phi_i$ of non-interacting model systems equals to the $\rho$-stationary 
wave functions at $\lambda = 0$ (i.e.$\Phi_i = \Psi^S_{\lambda=0}[\rho]$) and can be assigned to real 
electronic states $\Psi_j = \Psi[\rho,\nu,\alpha=1]$ \cite{gor2}. These assigned model states are the 
eigenstates of the GAC-KS formalism. As discussed above, they are eigenstates of a Hamiltonian operator 
with local multiplicative potential. In this way, the GAC will define the path of going from a non-
interacting system to an interacting system via a $\rho-$stationary path. Although for each of the 
interacting system, one can still end up with multiple non-interacting KS system. But with the criteria 
discussed previously it's possible to select the appropriate ones. So once the $\rho \Longleftrightarrow 
\hat v_{ext}$ for the interacting system is fixed, it do carries over to the KS system via GAC and vice 
versa. This shows how the proposed unified CS formalism not only establishes the density-to-potential 
mapping concretely but also constructs the KS system successfully. In the following sections we will 
exemplify what we have proposed so far through two model systems. This will be done in order address 
the critiques about density-to-potential mapping in $e$DFT.

\section{\lowercase{e}DFT beyond the HK and GL Theorem}
The issue of non-uniqueness in the density-to-potential mapping is also persuaded \cite{lpls} in 
the context of GL theorem \cite{gl1,gl2}. In \cite{lpls}, it has been demonstrated for  higher 
excited states of the considered $1D$ model system there is no equivalence of the GL/HK theorem. 
But the critical analysis of \cite{lpls} presented in this section will outline how the multiplicity 
of potentials still can't be ruled out even in the case of ground as well as lowest excited states. 
So one need to go beyond \cite{lpls} approach in order to address the validity of HK \& GL theorem 
for such state. In fact, relying on the principles of unified $e$DFT approach \cite{gor2,shjp2,
thesis,al} as discussed in Sec. II along with the proposed criteria of Sec. III, it will be shown 
here why the claim made in \cite{lpls} lacks merit to address the excited-state density-to-potential 
mapping. To validate the density-to-potential mapping (i.e. the analogue of HK/GL theorem) in 
\cite{lpls} proposed approach, we will consider as test cases: the examples of the $1D$ QHO with 
finite boundary and then the infinite potential well. 

For clarity in understanding let's first briefly discuss the theoretical formulation of \cite{lpls}. 
The Schr\"{o}dinger equation of two non-interacting fermions subjected to local one dimensional 
potentials $v(x)$ and $w(x)$ s.t. $v(x) \neq w(x) + C$, where $C$ is a constant are given by 
\begin{equation}
\Big[-\frac{1}{2}\frac{d^2}{dx^2} + v(x)\Big]\Phi_i(x) = \varepsilon_i \Phi_i(x)~,
\label{gl1}
\end{equation}
\begin{equation}
\Big[-\frac{1}{2}\frac{d^2}{dx^2} + w(x)\Big]\Psi_i(x) = \lambda_i \Psi_i(x)~.
\label{gl2}
\end{equation}
Suppose that the eigenfunctions of the local potential $w(x)$ generates the ground/excited-state 
eigendensity of $v(x)$ as one of it's eigendensity but with some arbitrary configuration which is
either same or different from the original one. Then one possible way of achieving this is: the 
wavefunctions $\Psi(x)$ of the potential $w(x)$ can be associated to the wavefunctions $\Phi(x)$ 
of the  potential $v(x)$ via the following unitary transformation i.e.
\begin{eqnarray}
\begin{pmatrix} 
\Psi_k(x) \\ \Psi_l(x)
\end{pmatrix}
&=&
\begin{pmatrix} \cos\theta(x) & \sin\theta(x) \\ -\sin\theta(x) & \cos\theta(x) \end{pmatrix}
\begin{pmatrix} \Phi_i(x) \\ \Phi_j(x) \end{pmatrix}\nonumber\\
&=&
\begin{pmatrix}
\Phi_i(x)\cos\theta(x)+\Phi_j(x)\sin\theta(x) \\-\Phi_i(x)\sin\theta(x)+\Phi_j(x)\cos\theta(x)
\end{pmatrix} ,
\label{gl3}
\end{eqnarray}
As a matter of which the density preserving constraint will be satisfied and the ground/excited state 
density of two potentials remain invariant i.e.
\begin{equation}
 \rho(x) = |\Phi_i(x)|^2 + |\Phi_j(x)|^2 = |\Psi_k(x)|^2 + |\Psi_l(x)|^2 .
 \label{gl4}
\end{equation}
Now the potentials can be obtained from the Eqs.(\ref{gl1}) and (\ref{gl2}) by inverting the same
\begin{eqnarray}
 v(x) &=& \varepsilon_i + \frac{\ddot{\Phi_i}(x)}{2\Phi_i(x)} = \varepsilon_j + \frac{\ddot{\Phi_j}(x)}
 {2\Phi_j(x)}\\
 w(x) &=& \lambda_k + \frac{\ddot{\Psi_k}(x)}{2\Psi_k(x)} = \lambda_l + \frac{\ddot{\Psi_l}(x)}{2\Psi_l(x)} ~.
 \label{gl5}
\end{eqnarray}
Also from Eqs.(\ref{gl1}) and (\ref{gl2}), the difference between any two eigenvalues $\Delta$ and $\Delta'$
corresponding to the potentials $v(x)$ and $w(x)$ are given by
\begin{equation}
\Delta = \varepsilon_j - \varepsilon_i = \frac{1}{2\Phi_i(x)\Phi_j(x)}\frac{d}{dx}[\Phi_j(x)
\dot{\Phi_i}(x) - \Phi_i(x)\dot{\Phi_j}(x)]~,
\label{gl6}
 \end{equation}
\begin{equation}
\Delta' = \lambda_k - \lambda_l = \frac{1}{2\Psi_k(x)\Psi_l(x)}\frac{d}{dx}[\Psi_l(x)\dot{\Psi_k}(x)
 - \Psi_k(x)\dot{\Psi_l}(x)]~.
\label{gl7}
\end{equation}
Now by plugging the values $\Psi_k(x)$ and $\Psi_l(x)$ from Eq.(\ref{gl3}) back in Eq.(\ref{gl7}), the
rotation $\theta(x)$ can be obtained from the following
\begin{equation}
\begin{array}{l}
\frac{d}{dx}[\dot{\theta}(x)\{\Phi^2_i(x)+\Phi^2_j(x)\}+\{\Phi_j(x)\dot{\Phi_i}(x)-\Phi_i(x)\dot
{\Phi_j}(x)\}]\\\\
=\Delta'[2\Phi_i(x)\Phi_j(x\cos2\theta(x)+\{\Phi^2_j(x)-\Phi^2_i(x)\}\sin2\theta(x)]
\label{gl8}
\end{array}
\end{equation}
or
\begin{equation}
 \rho(x)\ddot{\theta}(x)+\dot{\rho}(x)\dot{\theta}(x)+f(\Phi_i(x),\Phi_j(x),\Delta,\Delta',\theta)=0~,
\label{gl9}
 \end{equation}
where
\begin{eqnarray}
 f=&2\Delta\Phi_i(x)\Phi_j(x)-\Delta'[\Phi_i(x)\Phi_j(x)\cos2\theta(x)\nonumber\\
   &+\{\Phi^2_j(x)-\Phi^2_i(x)\}\sin2\theta(x)]~.
\label{gl10}
\end{eqnarray}
The Eq.(\ref{gl9}) is the central equation of \cite{lpls} theoretical framework which need to be solved 
numerically with proper initial conditions in order to obtain the alternate potential $w(x)$ for any
given density and eigenvalue differences. In this work, the adopted numerical procedure to solve the 
above mentioned differential equation is very much accurate even at the boundary where obtaining 
appropriate structure and behavior of the multiple potentials and the corresponding wavefunctions are
important and crucial.

\subsection{Results: $1D$ Quantum Harmonic Oscillator}
The first model system for demonstrating the density-to-potential mapping is the $1D$ QHO defined by 
\begin{eqnarray}
v(x) = \frac{1}{2}\omega^2x^2~,  {\mbox{where}}~~~-l\leq x \leq l ~.
\label{qhoeu1}
\end{eqnarray}
So the wavefunctions and energy eigenvalues of the $n^{th}$ eigenstate are given by
\begin{eqnarray}
\Phi_n(x)&=&\left(\frac{\omega}{\pi}\right)^\frac{1}{4}\frac{1}{\sqrt{2^n n!}}H_n(\sqrt{\omega}x)
\exp(-\frac{\omega x^2}{2})~,\label{eq28}\\
\varepsilon_n&=&(n+\frac{1}{2})\omega    ~,
\label{qhoeq2}
\end{eqnarray}
where $n=0,1,2.....$\\
(atomic units are adopted i.e. $\hbar = 1$ and $m_e = 1$)

\begin{figure}[h]
\begin{center}
\includegraphics[width=2.5in,height=3.0in,angle=0.0]{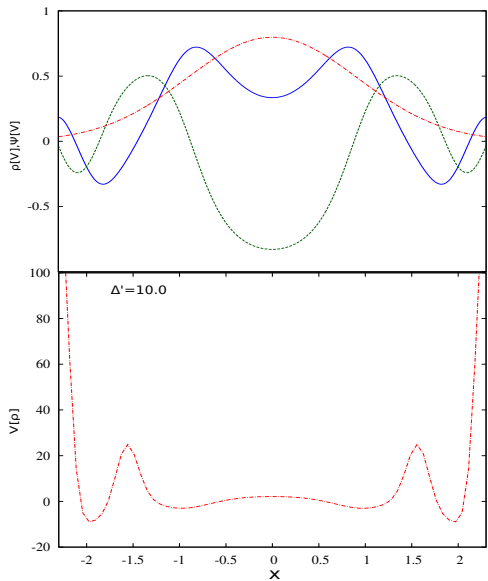} 
\end{center}
\caption{Upper panel: Shows (red color) the ground state density of the 1D QHO and the corresponding 
transformed wavefunctions $\Psi_k$ (blue) and $\Psi_l$ (green) for $\Delta' = 10.0$. Lower panel: 
Shows the alternate potential associated with above wavefunctions and density.}
\label{qhofig0_0-1}
\end{figure}

\begin{figure}[h]
\begin{center}
\includegraphics[width=2.5in,height=3.0in,angle=0.0]{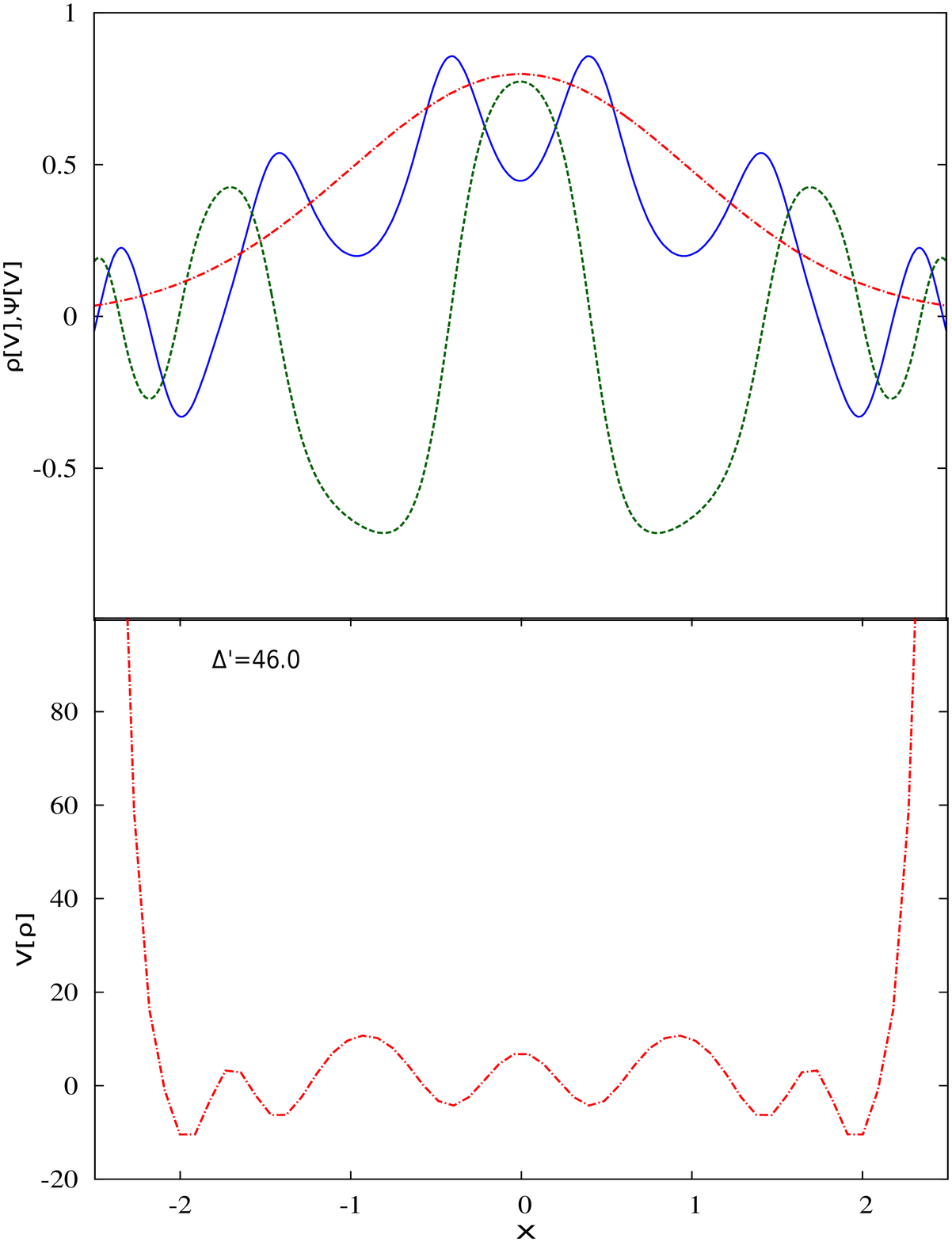} 
\end{center}
\caption{The figure caption is same as Fig.\ref{qhofig0_0-1} but with $\Delta' = 46.0$.}
\label{qhofig0_0-2}
\end{figure}

\subsection{Fermions in The Ground State}
Now consider two non-interacting fermions occupying the ground-state of the QHO i.e. $n = 0 = m$. So
the eigenvalue difference for this state $\Delta = \varepsilon_0 - \varepsilon_0 = 0$ and the density 
is given by
\begin{equation}
 \rho(x) = 2\left(\frac{\omega}{\pi}\right)^\frac{1}{2}\exp(-\omega x^2)~.
 \label{qhoeq3}
\end{equation}
Thus the corresponding equation for rotation $\theta(x)$ can be obtained from the Eq.(\ref{gl9}) and
is given by
\begin{equation}
 \rho(x)\ddot{\theta}(x) + \dot{\rho}(x)\dot{\theta}(x) - \Delta'[2\left(\frac{\omega}{\pi}\right)^\frac{1}{2}
 \exp(-\omega x^2)\cos2\theta(x)] = 0~.
 \label{qhoeq4}
\end{equation}
Now Eq.(\ref{qhoeq4}) has to be solved with proper initial conditions. The initial conditions can be fixed 
by taking into consideration the symmetry of the differential Eq.(\ref{qhoeq4}) and the normalization 
condition of the wavefunction. From Eq.(\ref{qhoeq4}) it is clear that $\frac{d\theta}{dx}|_{(x=0)} = 0$ 
as both $\Phi(x)$ and $\rho(x)$ are symmetric about $x=0$. Now another condition is that $\Psi_k(x)$ and 
$\Psi_l(x)$ must also be normalized. So if we plot the renormalization $R$ 
\begin{equation}
 \int^{l}_{-l} |\Psi_{k,l}(x)|^2 dx - 1 = R =0
 \label{qhoeq5}
\end{equation}
as a function of $\theta(x=0)$, then the points where $R = 0$ corresponds to the normalization of 
$\Psi_k(x)$ and $\Psi_l(x)$ \cite{lpls} and it will provide the initial condition on $\theta(x=0)$.
After finding $\theta(x)$, the transformed set of normalized wavefunctions $\Psi_k(x)$ and $\Psi_l(x)$
is being obtained. Again using these wavefunctions the potential $w(x)$ can be determined from the 
Eq.(\ref{gl5}). In Fig.\ref{qhofig0_0-1} and Fig.\ref{qhofig0_0-2}, we have shown two different 
potentials which are obtained for the eigenvalue differences $\Delta' = 10.0$ and $\Delta' = 46.0$ 
respectively along with the corresponding wavefunctions. The important point of observation here is 
that the ground-state density $\rho^{QHO} = \rho_0[v(x)=v_{QHO}(x),N=2]$ now corresponds to some arbitrary
excite-state having density $\rho_e[w(x) \neq v_{QHO}(x),N=2]$. As a matter of which, for the
fixed $\rho_0^{QHO}$ and $\Delta^\prime$, the system gets transformed to some other system $w(x)$
for which $Q^S[\rho_e[w(x)]=\rho_0^{QHO}]$ and/or $F[\rho_e[w(x)]=\rho_0^{QHO},\tilde{\rho_0}]$ will
be stationary. The corresponding stationary states are basically the transformed wavefunctions
which are given by Eq.(\ref{gl2}) $\Psi_l^S[w(x)] = \Psi_l[\rho_e,\tilde{\rho_0}]$. In this 
$\tilde{\rho_0}$ is the ground state density of the newly generated potential $w(x)$ and 
$\tilde{\rho_0} \neq \rho_0^{QHO}$. Now from the proposed criteria it follows that $\Delta T \neq 0$
, $\Delta[\rho_0(x),\tilde{\rho_0}(x)] > 0$ and new system is energetically far off from the original 
one. Hence the given ground-state density should be uniquely mapped to the original QHO potential 
$v(x)$ although there exist several multiple potentials $w(x)$. This result is consistent with the 
generalized/unified CS formalism \cite{gor2,shjp2,thesis,al}. 
\begin{figure}[h]
\begin{center}
\includegraphics[width=2.5in,height=3.0in,angle=0.0]{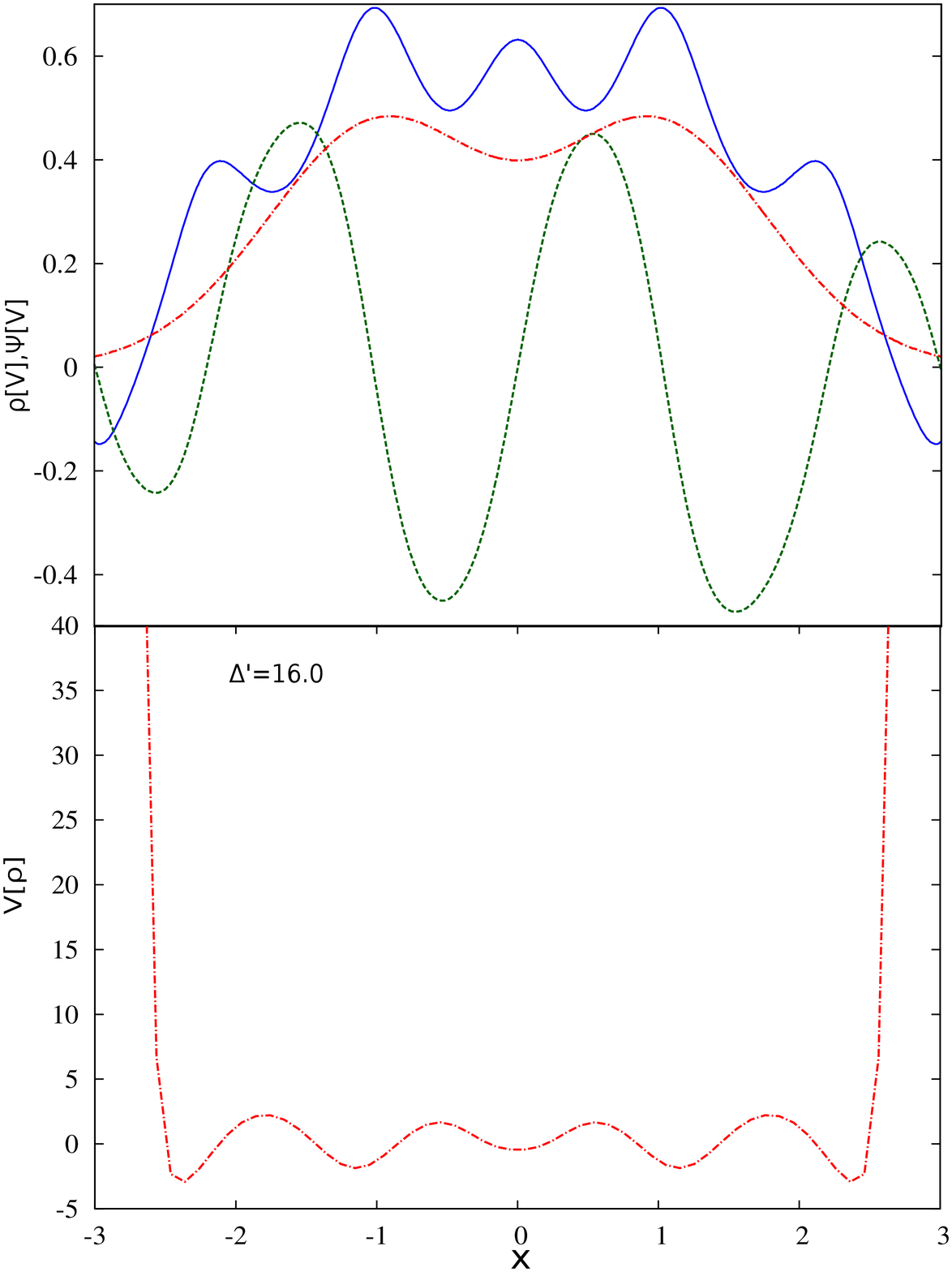} 
\end{center}
\caption{The figure caption is same as Fig.\ref{qhofig0_0-1} but for the lowest excited state density 
being produced with $\Delta' = 15.0$.}
\label{qhofig0_1-1}
\end{figure}

\begin{figure}[h]
\begin{center}
\includegraphics[width=2.5in,height=3.0in,angle=0.0]{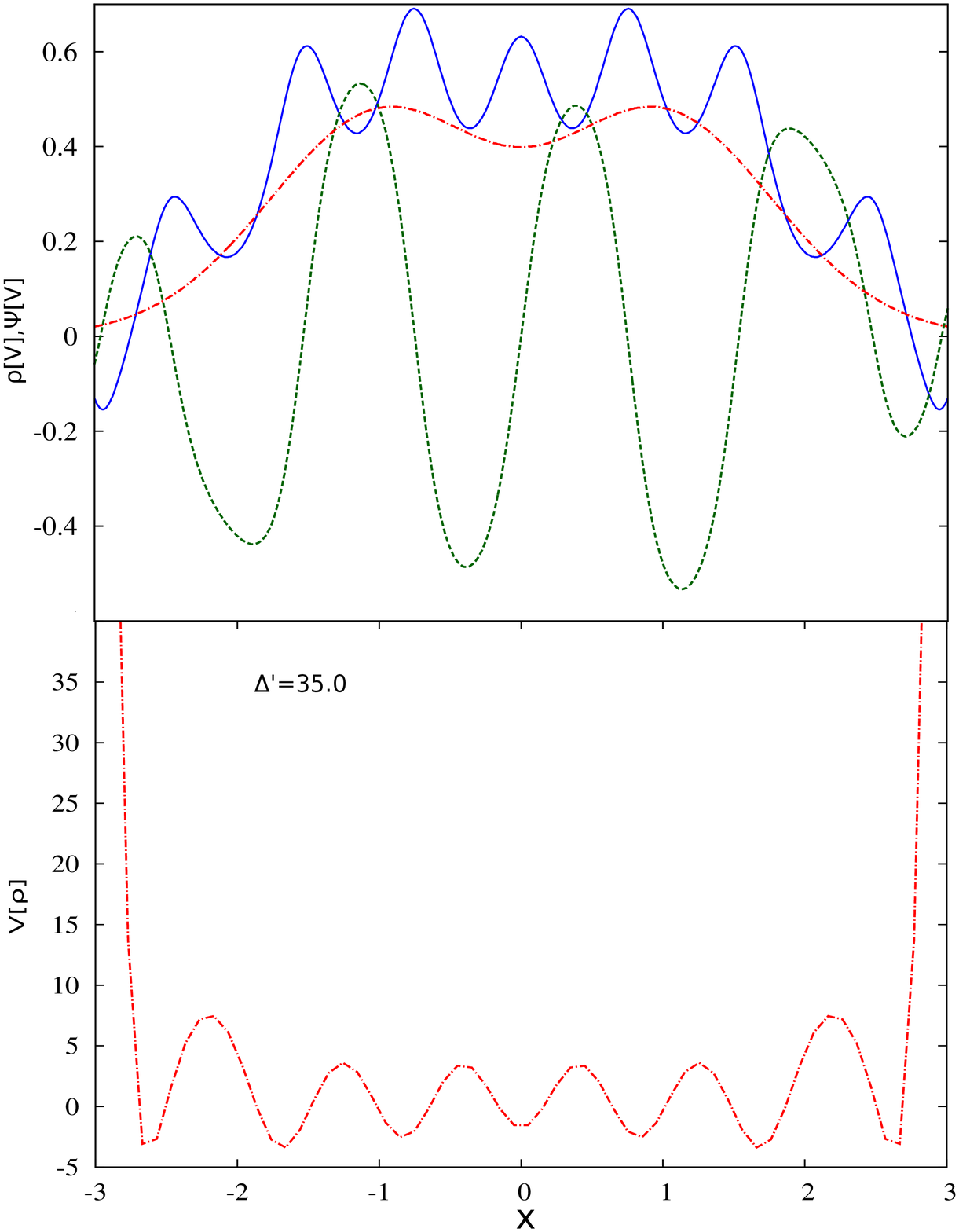} 
\end{center}
\caption{The figure caption is same as Fig.\ref{qhofig0_1-1} but with $\Delta' = 35.0$.}
\label{qhofig0_1-2}
\end{figure}

\subsection{Fermions in The Lowest Excited State}
As the second example, we consider the lowest excited-state of the QHO. So the two non-interacting 
fermions are now occupying the $n=0$ and $m=1$ state. For this case, $\varepsilon_0 = \frac{1}{2}\omega$,
$\varepsilon_1 = \frac{3}{2}\omega$ and $\Delta = \varepsilon_1 - \varepsilon_0 = \omega$ with the 
density
\begin{equation}
\rho(x) = \left(\frac{\omega}{\pi}\right)^\frac{1}{2}\exp(-\omega x^2)(1 + 2\omega x^2),
\label{qhoeq6}
\end{equation}
and the corresponding equation for rotation $\theta(x)$ is given by
\begin{eqnarray}
\rho(x)\ddot{\theta}(x) + \dot{\rho}(x)\dot{\theta}(x) + 2\omega x\left(\frac{2\omega^2}{\pi}\right)^
\frac{1}{2}\exp(-\omega x^2)\nonumber\\ 
-\Delta'[2x\left(\frac{2\omega^2}{\pi}\right)^\frac{1}{2}\exp(-\omega x^2)\cos2\theta(x) + \nonumber\\
\frac{\omega}{\pi}\exp(-\omega x^2)\{2\omega x^2 - 1\}\sin2\theta(x)] = 0~.
 \label{eq34}
\end{eqnarray}
Since in this case $\Phi_0(x)$ is symmetric, $\Phi_1(x)$ is antisymmetric, so $\rho(x)$ symmetric around 
$x=0$. Thus Eq.(\ref{eq34}) implies that $\theta(x)$ should be symmetric at $x=0$. The initial conditions 
on $\frac{d\theta}{dx}|_{(x=0)}$ is obtained from the behavior of the renormalization $R$ as a function 
of $\frac{d\theta}{dx}|_{(x=0)}$. Following the same procedure as before, in this case also we have obtained 
different potentials for the fixed lowest excited state density which are shown in the  Fig.\ref{qhofig0_1-1} 
and Fig.\ref{qhofig0_1-2}. These two alternative potentials and the transformed wavefunctions correspond 
to two different eigenvalue differences $\Delta' = 16.0$ and $\Delta' = 35.0$. As described in the ground 
state case, in this case also the structure of the  potential is different from the original 1D QHO as the 
potential should follow the structure of the wavefunctions. However, according to the unified CS $e$DFT, 
the results are never due to the violation of GL theorem. This is because the ground and lowest excited 
states of the newly found potential are quite different from that of the QHO. So following similar argument 
as in the previous case, now the lowest excited-state density of the QHO corresponds to some different 
eigendensity of the multiple potentials. Thus the multitude of potentials poses no issues for the validity 
of the GL theorem.

\begin{figure}[h]
\begin{center}
\includegraphics[width=2.5in,height=3.0in,angle=0.0]{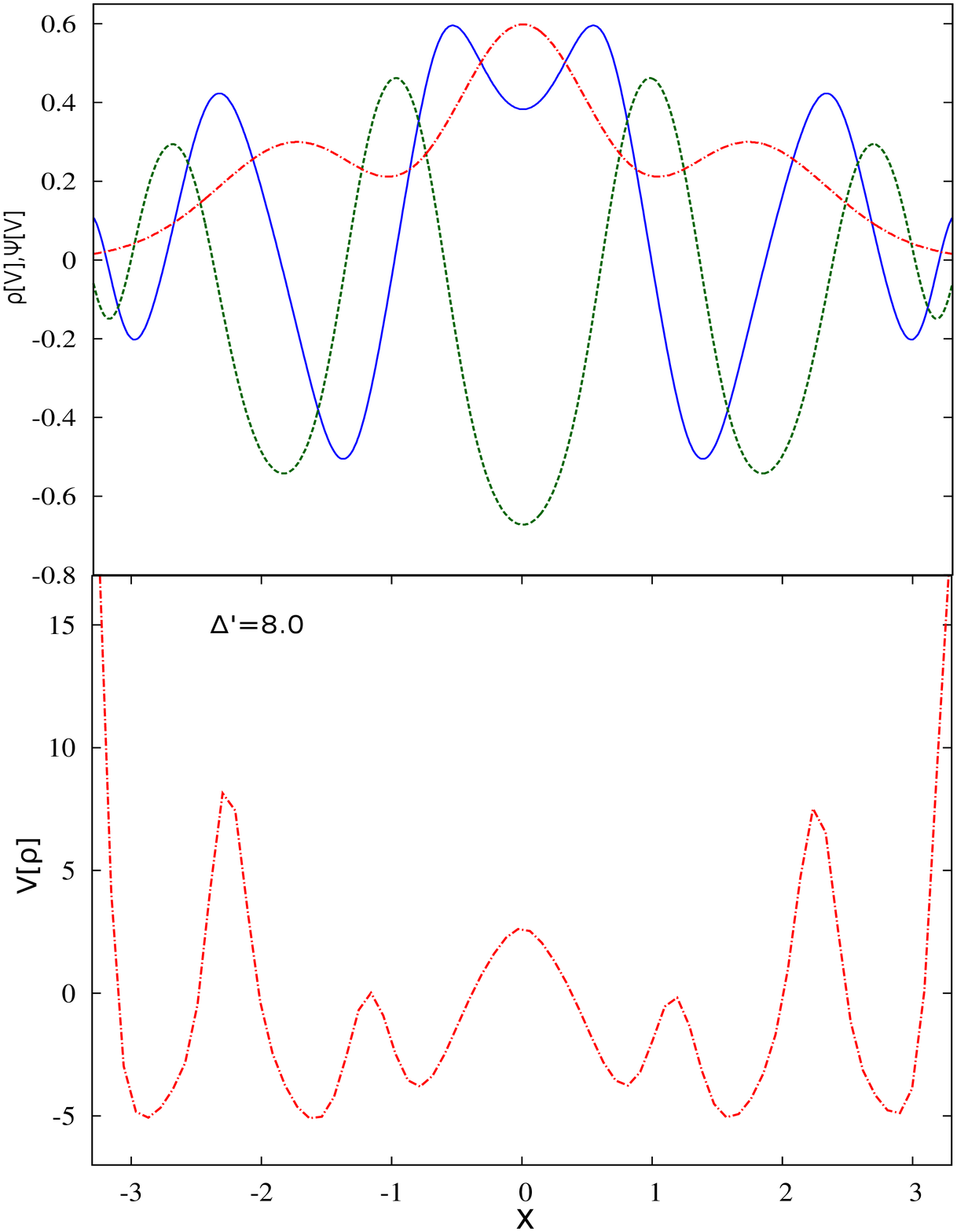}
\end{center}
\caption{The figure caption is same as Fig.\ref{qhofig0_0-1} but for one of the higher excited state 
density being produced with $\Delta' = 8.0$.}
\label{qhofig0_2-1}
\end{figure}

\begin{figure}[h]
\begin{center}
\includegraphics[width=2.5in,height=3.0in,angle=0.0]{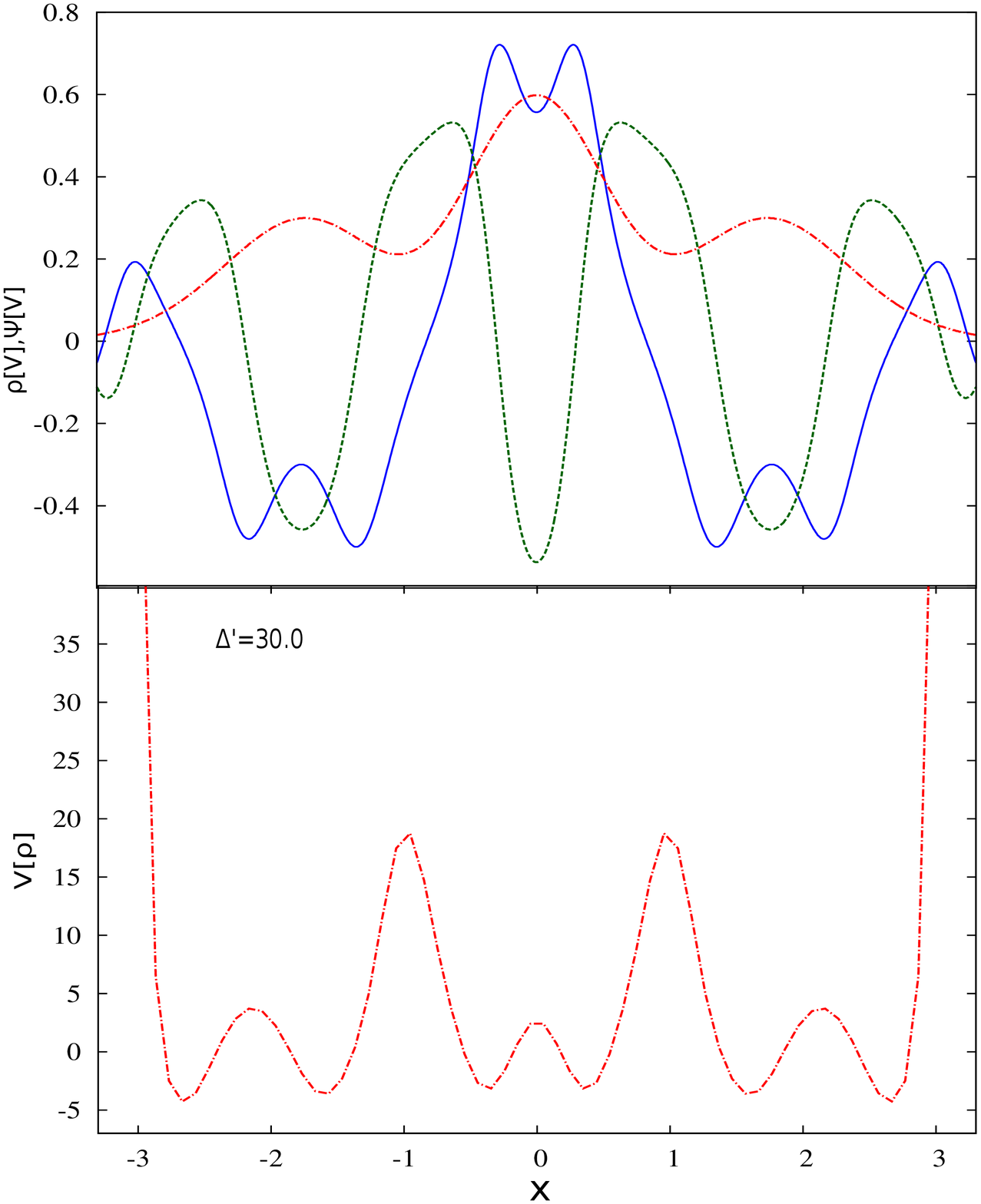}
\end{center}
\caption{The figure caption is same as Fig.\ref{qhofig0_2-1} but with $\Delta' = 30.0$.}
\label{qhofig0_2-2}
\end{figure}

\begin{figure}[h]
\begin{center}
\includegraphics[width=3.0in,height=3.5in,angle=0.0]{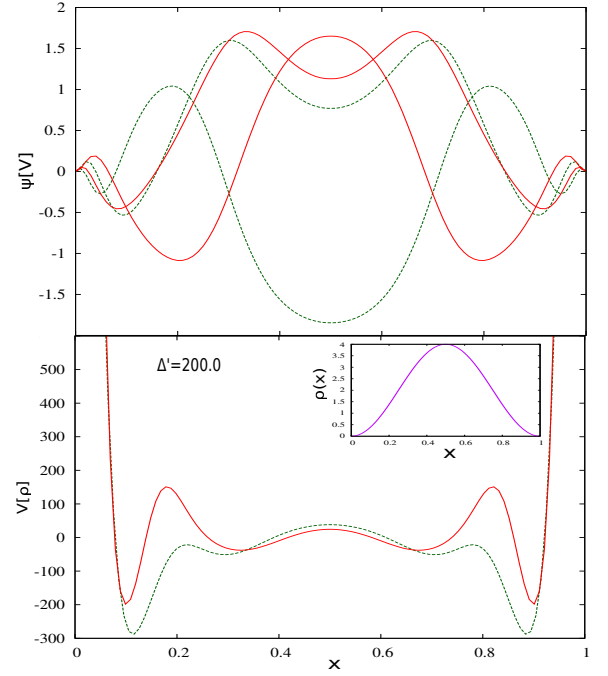} 
\end{center}
\caption{Upper panel: Shows the alternate wavefunctions $\Psi_k$ and $\Psi_l$ (green \& red) resulting the
ground state density of 1D potential well for $\Delta' = 200.0$. Lower panel: Shows the alternate potentials 
(green \& red) and the density (magenta) associated with above wavefunctions.}
\label{pbfig2-1}
\end{figure}

\begin{figure}[h]
\begin{center}
\includegraphics[width=3.0in,height=3.5in,angle=0.0]{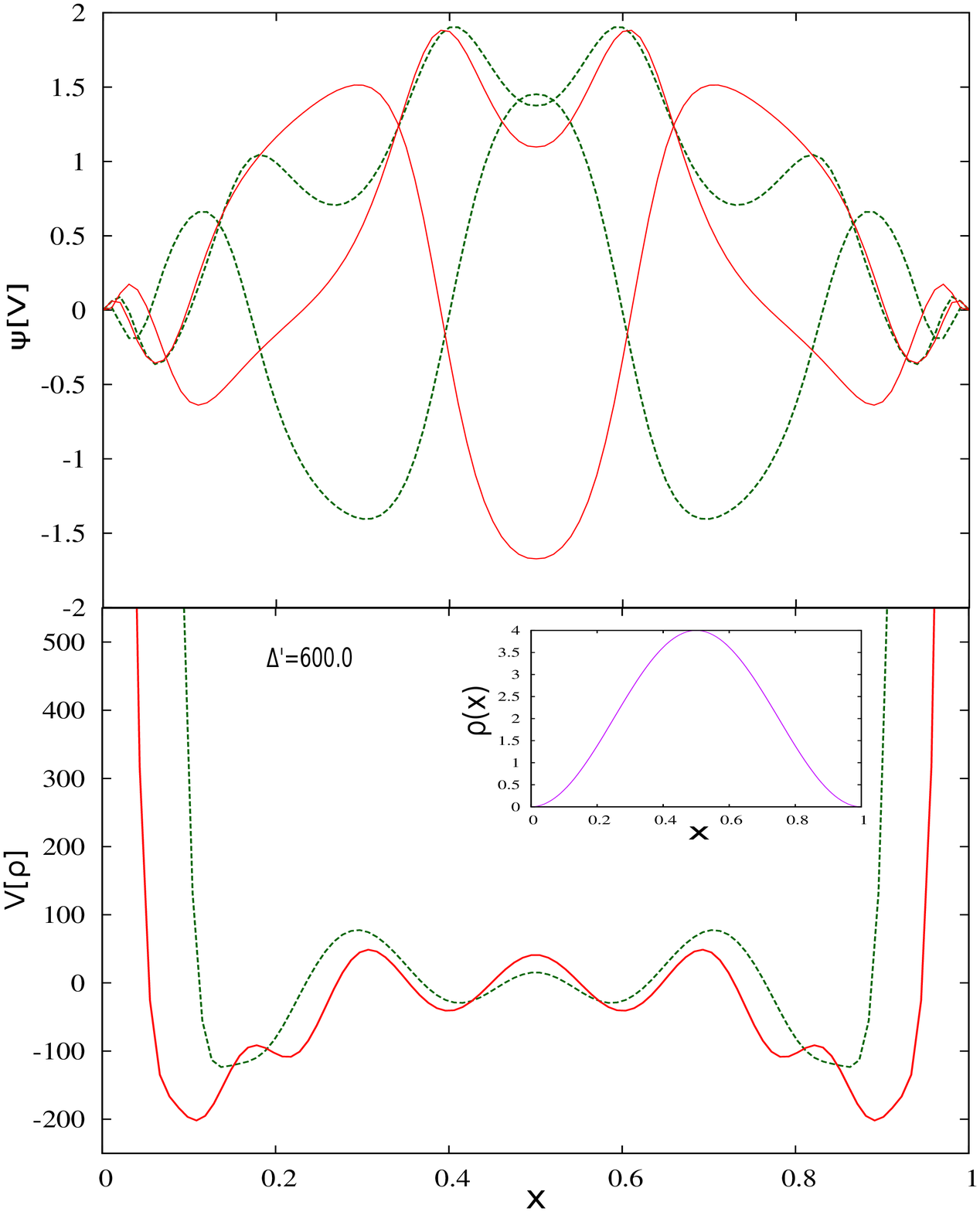} 
\end{center}
\caption{The figure caption is same as Fig.\ref{pbfig2-1} but with $\Delta' = 600.0$.}
\label{pbfig2-2}
\end{figure}

\begin{figure}[h]
\begin{center}
\includegraphics[width=3.0in,height=3.5in,angle=0.0]{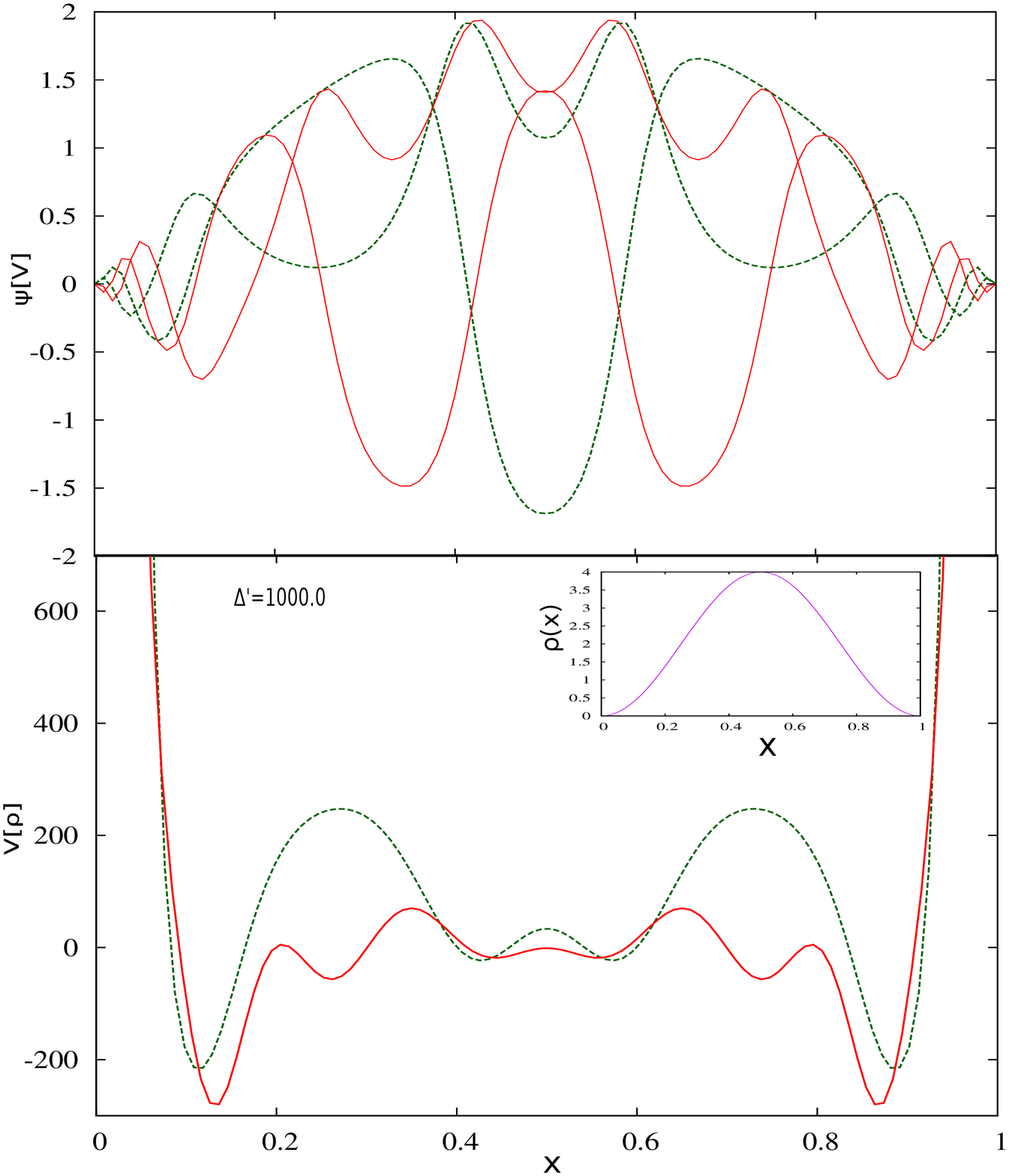} 
\end{center}
\caption{The figure caption is same as Fig.\ref{pbfig2-1} but with $\Delta' = 1000.0$.}
\label{pbfig2-3}
\end{figure}

\begin{figure}[h]
\begin{center}
\includegraphics[width=3.0in,height=3.5in,angle=0.0]{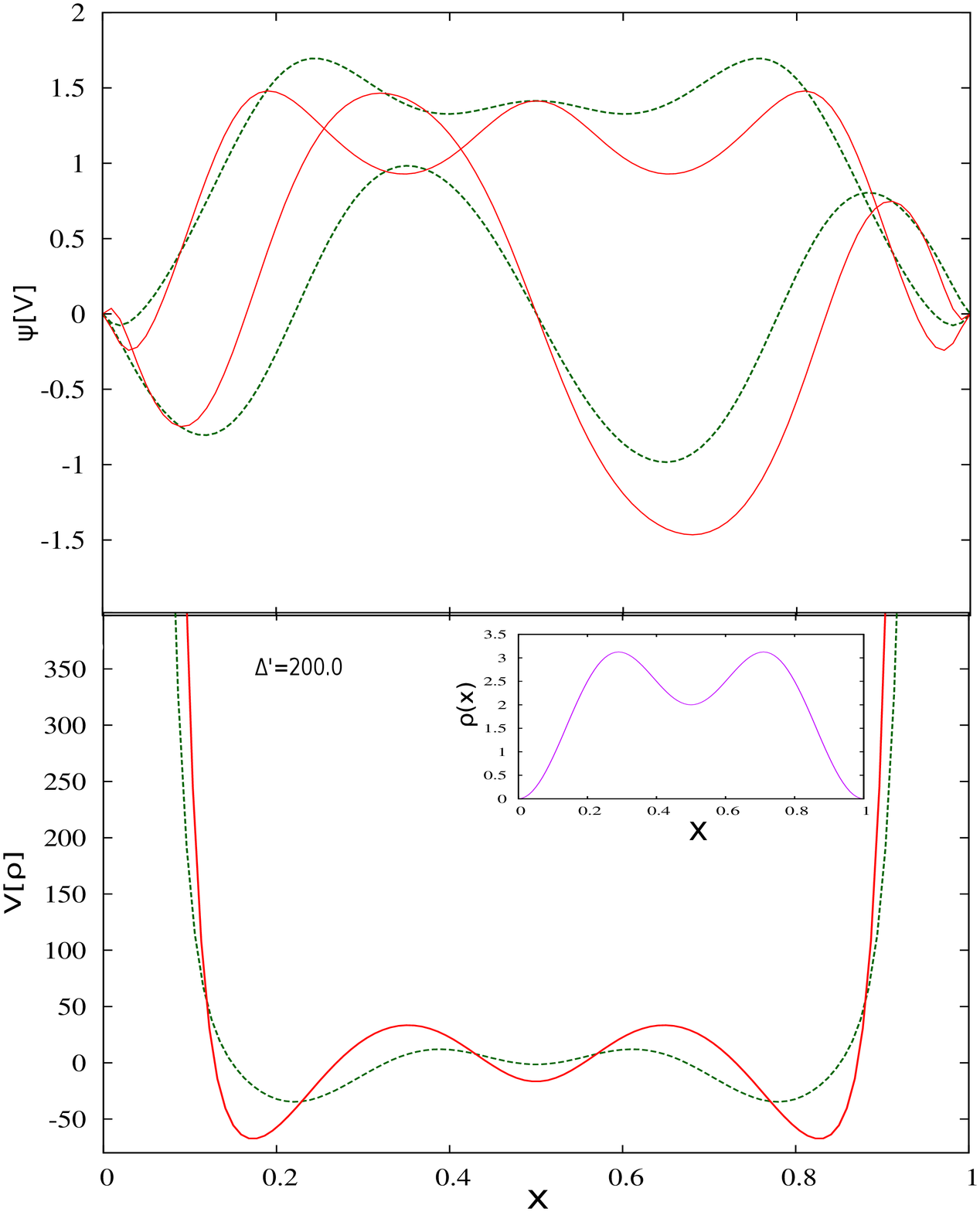} 
\end{center}
\caption{The figure caption is same as Fig.\ref{pbfig2-1} but for the lowest excited state density
being produced with $\Delta' = 200.0$.}
\label{pbfig3-1}
\end{figure}

\begin{figure}[h]
\begin{center}
\includegraphics[width=3.0in,height=3.5in,angle=0.0]{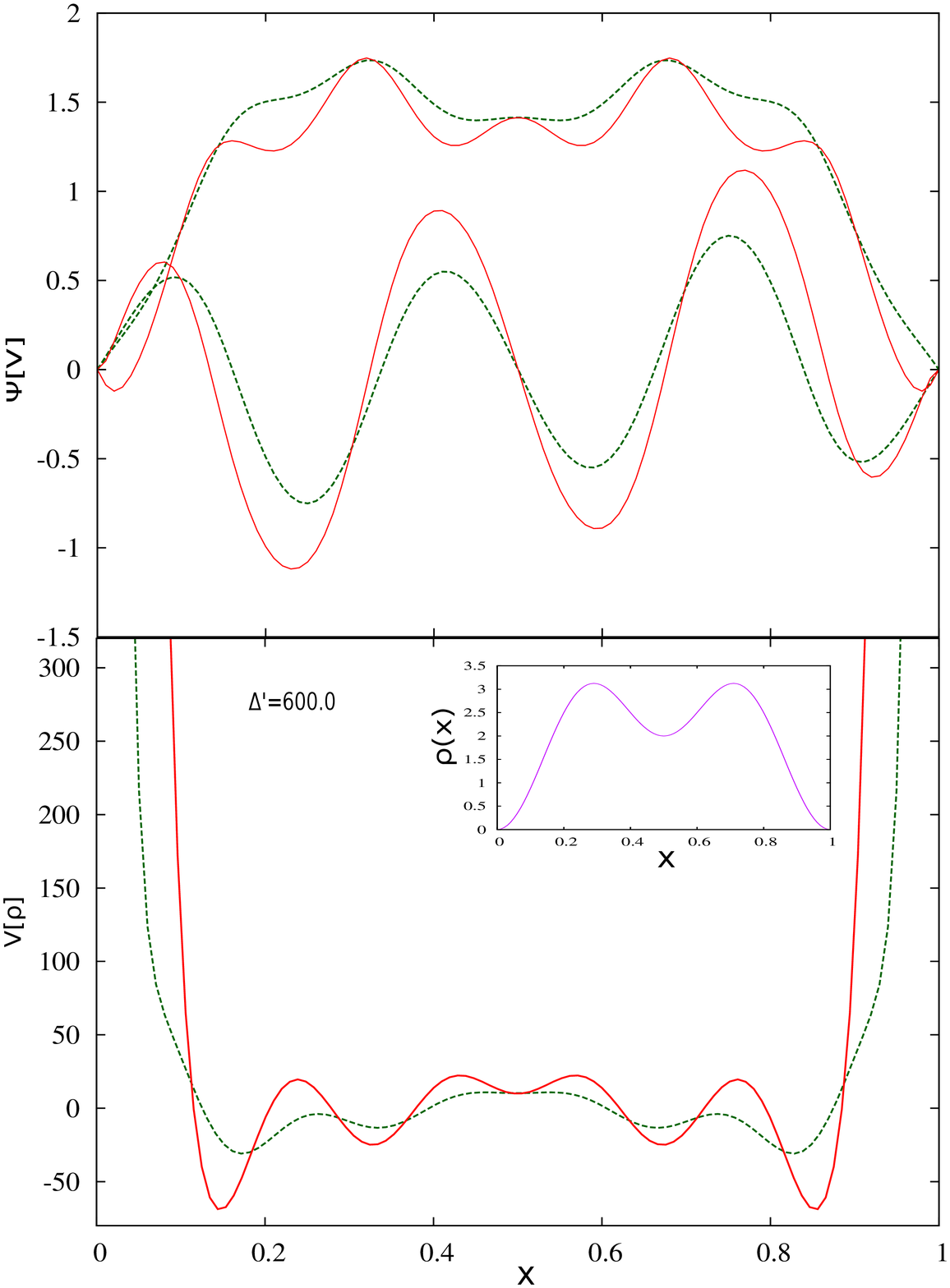} 
\end{center}
\caption{The figure caption is same as Fig.\ref{pbfig3-1} but with $\Delta' = 600.0$.}
\label{pbfig3-2}
\end{figure}

\begin{figure}[h]
\begin{center}
\includegraphics[width=3.0in,height=3.5in,angle=0.0]{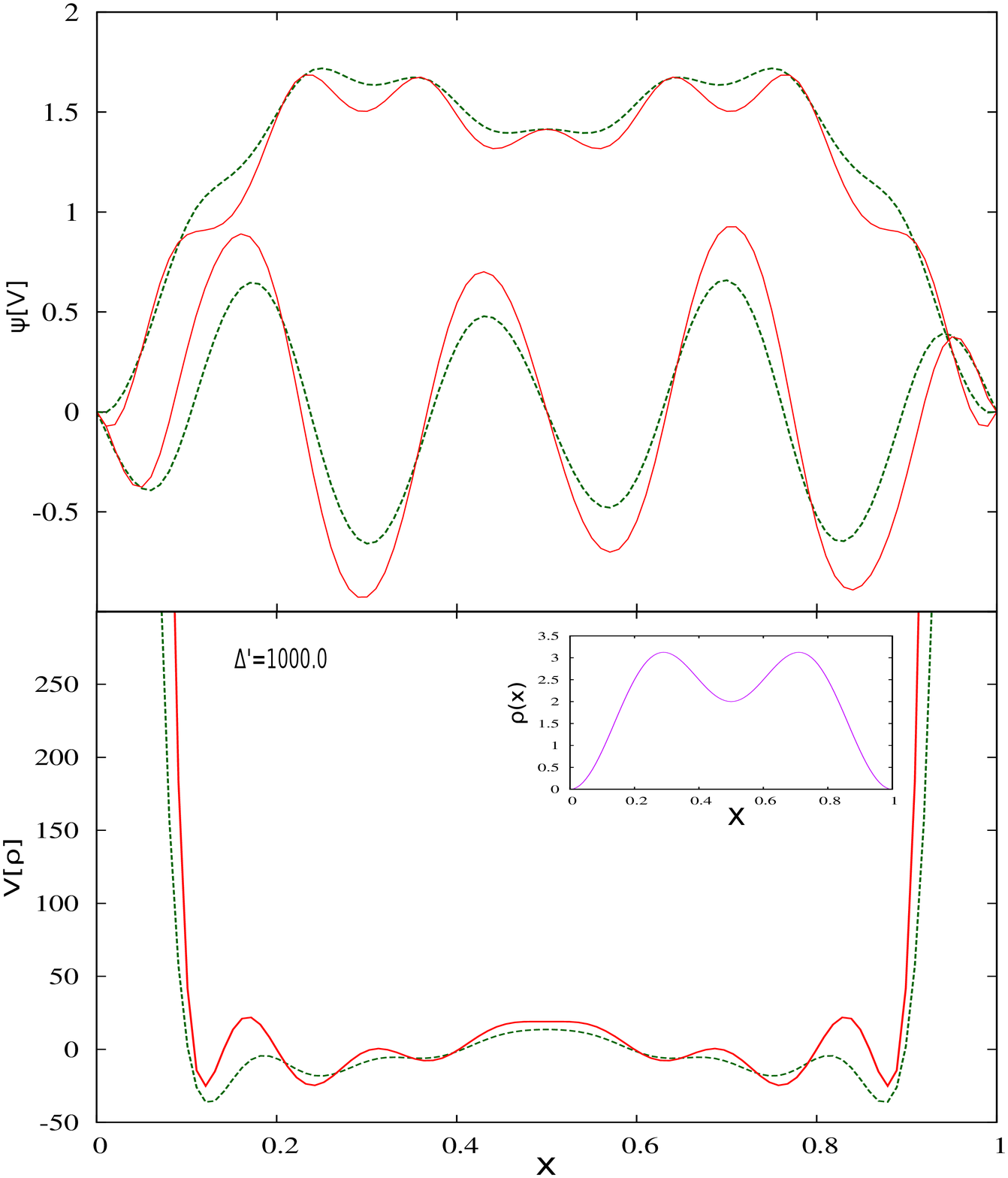} 
\end{center}
\caption{The figure caption is same as Fig.\ref{pbfig3-1} but with $\Delta' = 1000.0$.}
\label{pbfig3-3}
\end{figure}

\begin{figure}[h]
\begin{center}
\includegraphics[width=3.0in,height=3.5in,angle=0.0]{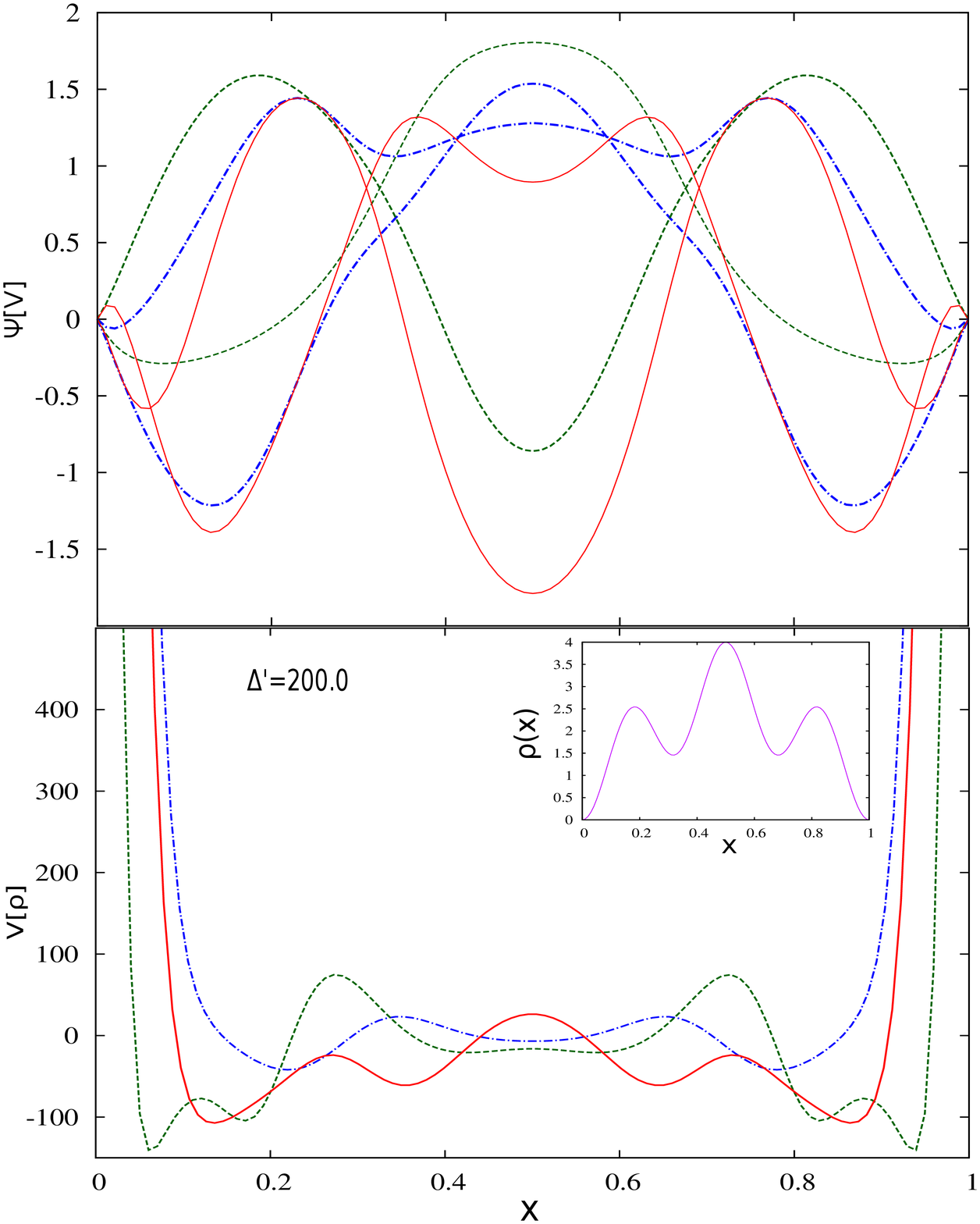} 
\end{center}
\caption{The figure caption is same as Fig.\ref{pbfig2-1} but for one of the higher excited state density
being produced with $\Delta' = 200.0$.}
\label{pbfig4-1}
\end{figure}

\begin{figure}[h]
\begin{center}
\includegraphics[width=3.0in,height=3.5in,angle=0.0]{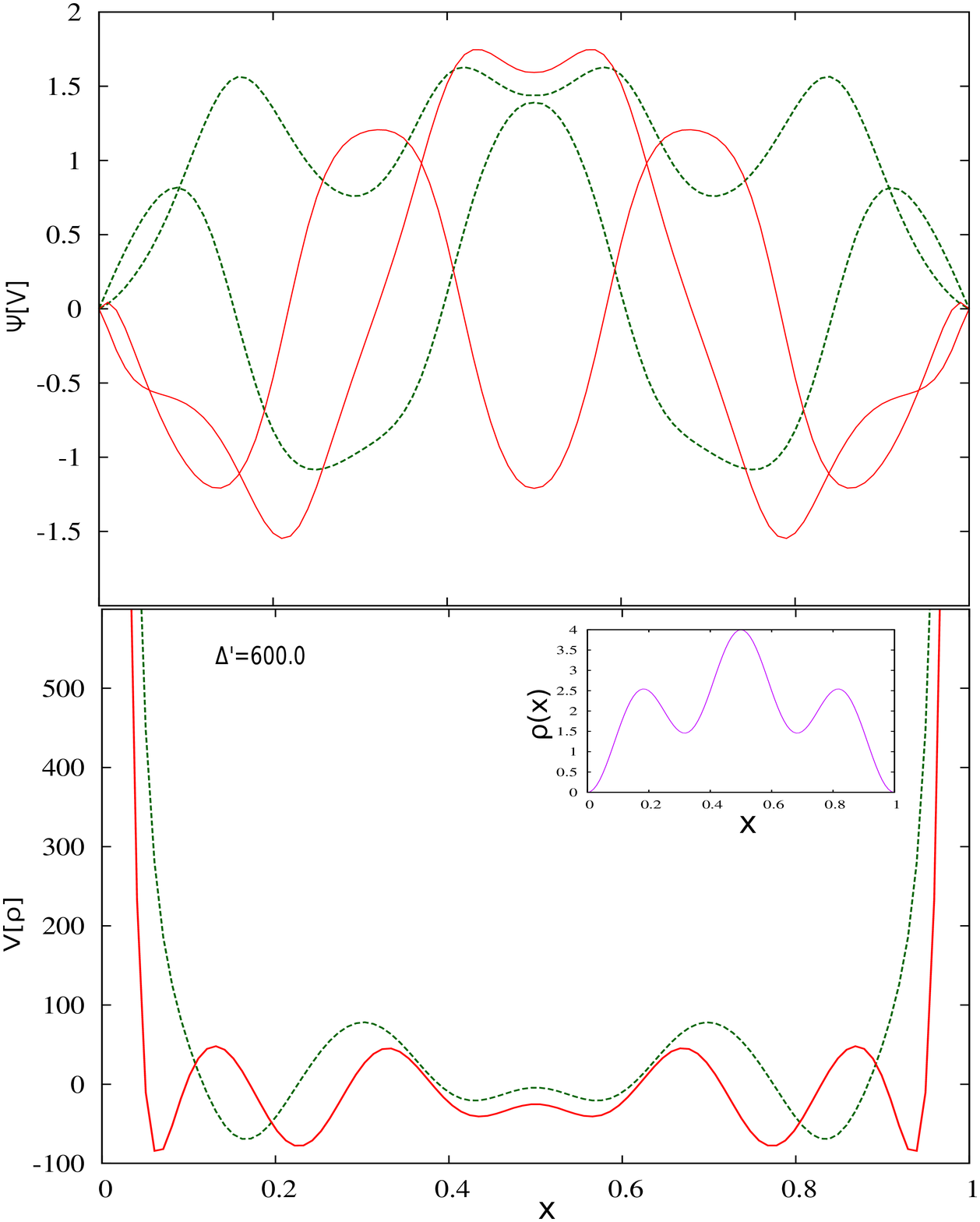} 
\end{center}
\caption{The figure caption is same as Fig.\ref{pbfig4-1} but with $\Delta' = 600.0$.}
\label{pbfig4-2}
\end{figure}

\begin{figure}[h]
\begin{center}
\includegraphics[width=3.0in,height=3.5in,angle=0.0]{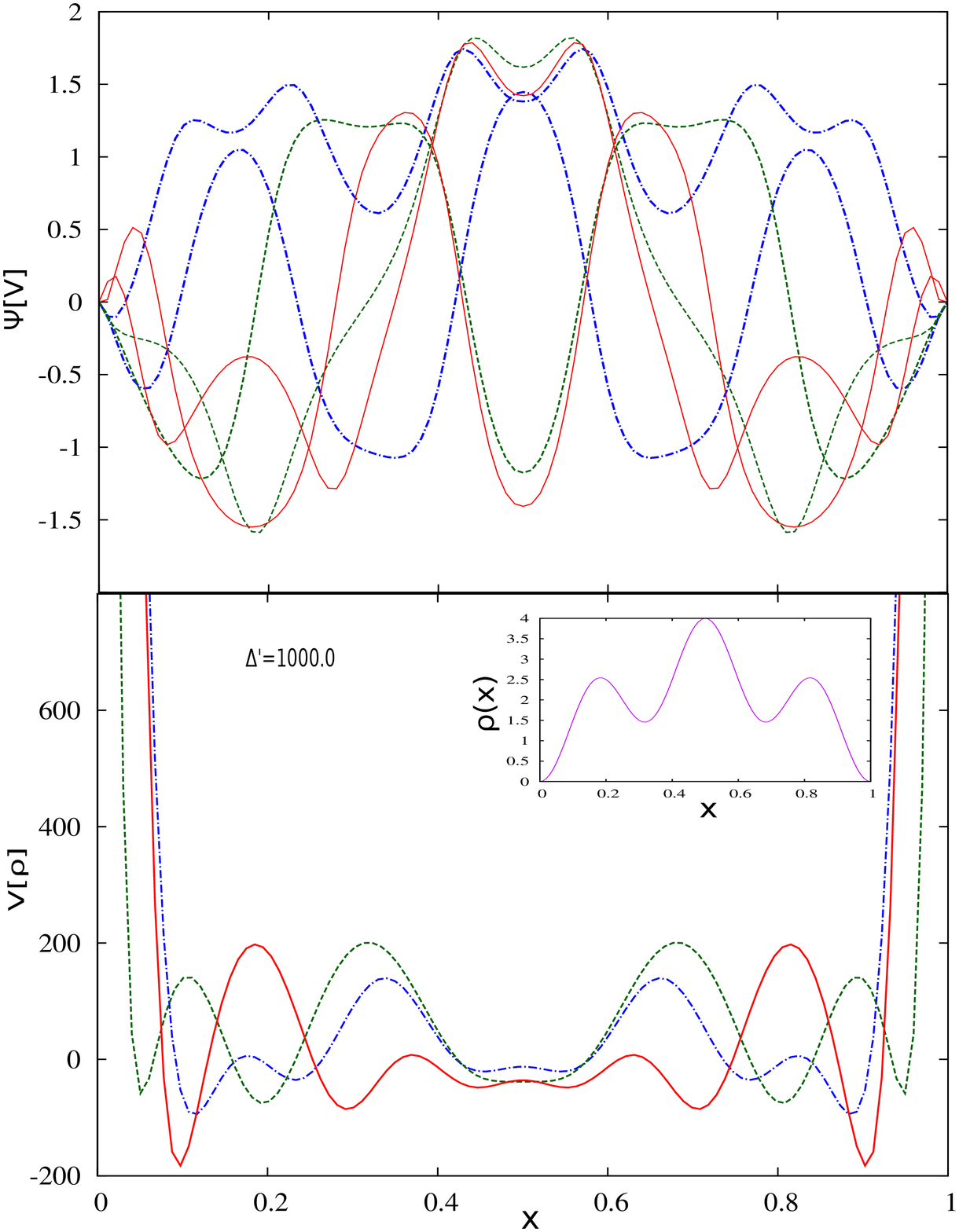} 
\end{center}
\caption{The figure caption is same as Fig.\ref{pbfig4-1} but with $\Delta' = 1000.0$.}
\label{pbfig4-3}
\end{figure}

\subsection{Fermions in Higher Excited States}
Here we consider one of the higher excited-state of 1D QHO (i.e. two non-interacting fermions are in the 
$n=0$ and $m=2$ states). For this case, the eigenvalue difference is $\Delta = \varepsilon_2 - \varepsilon_0 
= 2\omega$ and the density corresponding to it is given by
\begin{equation}
\rho(x) = \left(\frac{\omega}{\pi}\right)^{\frac{1}{2}}\exp(-\omega x^2)\{1 + (1 - 2\omega x^2)^2\}~.
\label{qho02-1}
\end{equation}
Similarly, the corresponding equation for rotation $\theta(x)$ is given by
\begin{eqnarray}
\rho(x)\ddot{\theta}(x) + \dot{\theta}(x)\dot{\theta}(x) + 4\omega\left(\frac{\omega}{2\pi}\right)^
{\frac{1}{2}}(2\omega x^2 -1)\exp(-\omega x^2) \nonumber\\
-\Delta'[\left(\frac{\omega}{2\pi}\right)^{\frac{1}{2}}(2\omega x^2 -1)\exp(-\omega x^2) \cos2\theta(x)
\nonumber\\
+ \left(\frac{\omega}{\pi}\right)^{\frac{1}{2}}\exp(-\omega x^2)\{\frac{1}{2}(2\omega x^2 - 1)^2 - 1\}
\sin2\theta(x)] = 0~.\nonumber\\
\label{qho02-2}
\end{eqnarray}
Now by solving Eq.(\ref{qho02-2}) for rotation $\theta(x)$ in analogous with the ground-state of the QHO
and after taking care of the normalization of the transformed wavefunctions, the potential $w(x)$ is 
obtained for $\Delta' = 8.0,~30.0$. The potentials along with the wavefunctions are shown in Fig.
\ref{qhofig0_2-1} \& Fig.\ref{qhofig0_2-2}. Similar to ground and lowest excited-state, here too the 
given density is a different eigendensity of the new potentials. If it would have the same eigendensity
of $w(x)$ then $w(x)$ should have been identical to the $v_{QHO}(x)$. But it is not the case. That's 
why the generated potentials are completely different from the QHO.

\subsection{Results: $1D$ Infinite Potential Well}
As our second case study, we consider the model system same as that reported in \cite{lpls} (i.e. particles 
are trapped inside an $1D$ infinite potential well). For an infinite potential well with length varying from 
$0$ to $1$, the $n^{th}$ eigenfunction $\Phi_n(x)$  and the energy eigenvalue $\varepsilon_n$ are given by
\begin{equation}
\Phi_n(x) = \sqrt{2}\sin (n\pi x) ~;~ \varepsilon_n = \frac{n^2\pi^2}{2}~,
\label{pbeq1}
\end{equation}
where $n=1, 2, 3 ....$. The density $\rho(x)$ corresponding to the two potentials $v(x)$ and $w(x)$
is given by Eq.(\ref{gl4}).

\subsection{Fermions in The Ground State}
For two spinless non-interacting particles in $n=1=m$ states, the energies of two states and the 
difference are 
\begin{equation}
\varepsilon_1 = \frac{\pi^2}{2} = \varepsilon_2 ~;~ \Delta = \varepsilon_2 - \varepsilon_1 = 0~.
\label{pbeq3}
\end{equation}
The density corresponding to these states is
\begin{equation}
\rho(x) = 4[\sin^2(\pi x)]~,
\label{pbeq5}
\end{equation}
and the equation corresponding to Eq.(\ref{gl9}) for the rotation $\theta(x)$ is 
\begin{equation}
\rho(x)\ddot{\theta}(x) + \dot{\rho}(x)\dot{\theta}(x) - \Delta'[4\sin^2\pi x \cos2\theta(x)] = 0~.
\label{pbeq6}
\end{equation}
Since $\Phi_1(x)$ is symmetric and $\rho(x)$ is symmetric about $x=\frac{1}{2}$. Thus Eq.(\ref{pbeq6})
indicates that $\theta(x)$ should be symmetric such that $\dot{\theta}(\frac{1}{2})=0$. With this initial 
condition and choosing any value of $\Delta'$ one can solve for $\theta(x)$  and subsequently obtain the 
$\Psi_k$~s. Now using these $\Psi_k$~s, the alternate potentials $w(x)$ will be obtained by using Eq.
(\ref{gl5}). Since the transformed wavefunction $\Psi_k(x)$ must also be normalized. This condition will 
be fulfilled by choosing the appropriate value of $\theta(\frac{1}{2})$ at which the $\Psi_k(x)$ should 
be normalized. Once $\Psi_k(x)$ is normalized then $\Psi_l(x)$ will also be normalized. Again by adopting 
the same procedure as that described in the case of $1D$ QHO, the alternative multiple potentials are
obtained by making use of the following renormalization $R$ condition   
\begin{equation}
 \int^1_0 |\Psi_k(x)|^2 dx - 1 = R = 0~.
 \label{pbeq7}
\end{equation}
All the wavefunctions, densities and multiple potentials are shown in the Figs.(\ref{pbfig2-1} $to$ 
\ref{pbfig2-3}). Here we have generated the multiple potentials for $\Delta' = 200.0,~600.0~\&~1000.0$
respectively. As is expected, the wavefunctions are totally different from the ground state of the
$1D$ infinite well. Although the density remains to be the same in all the cases. But its not the
ground state eigendensity of the multiple potentials. So this poses no issue for the HK theorem.

\subsection{Fermions in The Lowest Excited State}
Now consider two fermions occupying the $n=1,~m=2$ (i.e. the lowest excited-state) eigenstates of the 
infinite potential well. Here too we have obtained several multiple potentials unlike \cite{lpls}. For 
this excited-state, the energy eigenvalues are $\varepsilon_1 = \frac{\pi^2}{2}$, $\varepsilon_2 = 
2\pi^2$ with $\Delta = \frac{3\pi^2}{2}$. Hence the density arising from these two states is given by 
\begin{equation}
\rho(x) = 2[\sin^2(\pi x) + \sin^2(2\pi x)]~.
\label{pbeq8}
\end{equation}
Similar to the previous examples, the equation for the rotation $\theta(x)$ is the following
\begin{eqnarray}
 \rho(x)\ddot{\theta}(x) + \dot{\rho}(x)\dot{\theta}(x) + 6\pi^2\sin(\pi x)\sin(2\pi x)\nonumber\\
 -\Delta'[4\sin(\pi x)\sin(2\pi x)\cos2\theta(x)\nonumber\\
 + 2\{\sin^2(2\pi x) - \sin^2(\pi x)\}\sin2\theta(x)] = 0~.
\label{pbeq9}
\end{eqnarray}
Here $\Phi_1(x)$ is symmetric, $\Phi_2(x)$ is antisymmetric and $\rho(x)$ symmetric about $x = \frac{1}
{2}$. Thus Eq.(\ref{pbeq9}) predicts that $\theta(x)$ is antisymmetric such that $\theta(\frac{1}{2}) = 0$. 
In this case also normalization of both $\Psi_k(x)$ and $\Psi_l(x)$ are taken care and the proper $R$ 
(renormalization) values are obtained w.r.t. $\frac{d\theta}{dx}(\frac{1}{2})$. Quite interestingly, 
in this case also we have successfully generated multiple potentials for $\Delta' = 200.0,~600.0~\&~1000.0$. 
This is where \cite{lpls} failed to explain the validity of GL theorem. As expected, the potential 
follows the wavefunctions pattern. This is obvious at the boundary where the wavefunctions are 
perfectly vanishing, the potential shoots up to a very large positive value. The potentials along 
with wavefunctions are shown in the Figs.(\ref{pbfig3-1} $to$ \ref{pbfig3-3}). Following the same
argument as in the case of the previous model system, the multiplicity of potentials obtained here are
nothing to do with the GL theorem.

\subsection{Fermions in Higher Excited States}
Now to complete our exploration on $1D$ well, we have considered here the second excited-state of 
it. This is the only excited-state for which \cite{lpls} reported multiple external potentials for 
various eigenvalue differences. We too generated multiple potentials and the corresponding wavefunctions 
for $\Delta' = 200.0,~600.0~\&~1000.0$ which are shown in the Figs.(\ref{pbfig4-1} $to$ \ref{pbfig4-3}). 
The results follow the trend similar to that of the ground and lowest excited-state. In all the cases, 
we have noticed that the potentials and the corresponding rotation angles can never attain flat 
structure at the boundary unlike \cite{lpls}.

\begin{figure}[h]
\begin{center}
\includegraphics[width=3.3in,height=3.0in,angle=0.0]{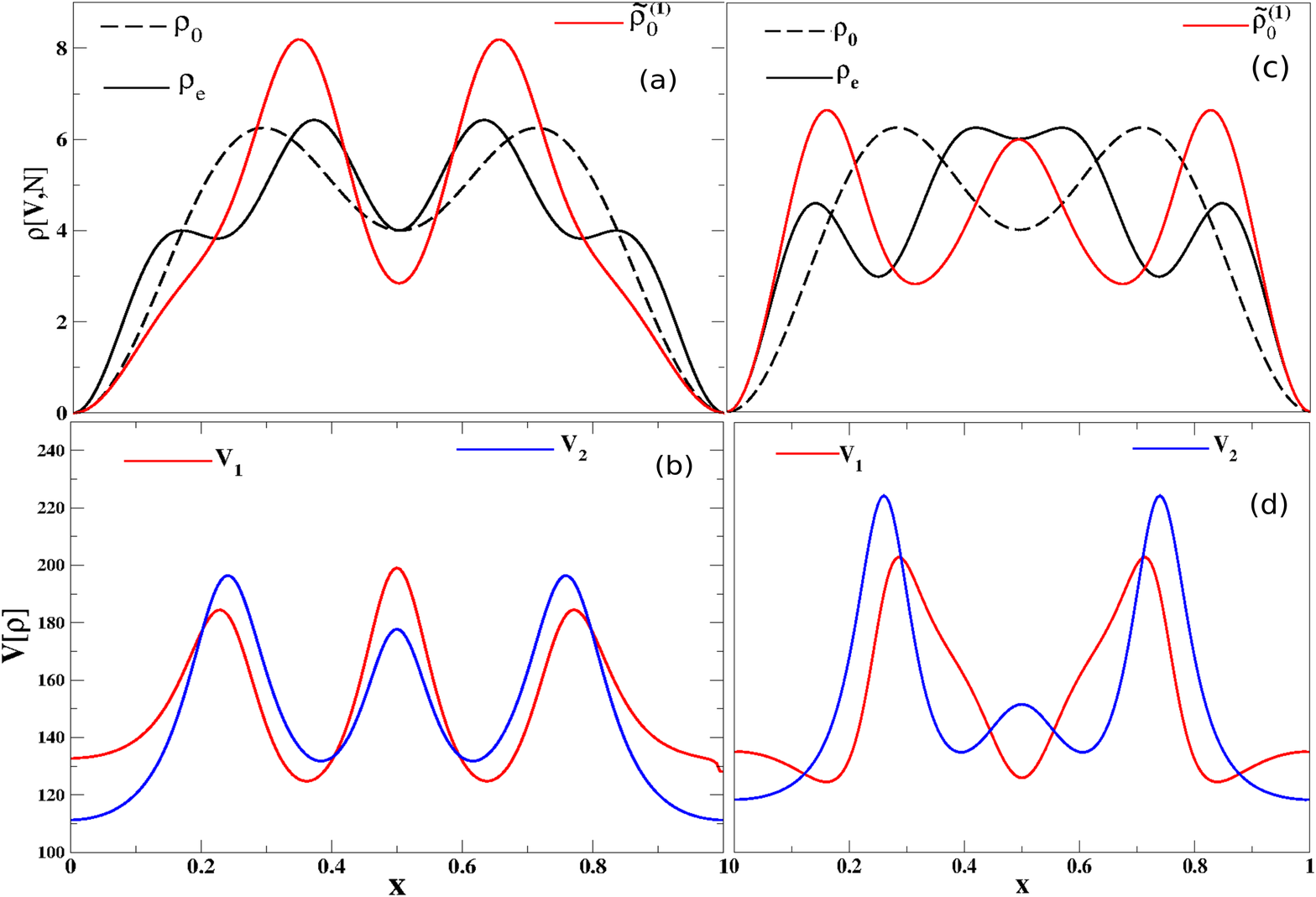} 
\end{center}
\caption{(a) $\rho_e[n_1(2),n_2(1),n_4(1)]$ is the excited-state density of $1D$ potential well 
with ground-state $\rho_0 $. $\tilde{\rho}^{(1)}_0$ is the ground-state density of potential 
$V_1$ whose excited-state configuration $[n_1(2),n_2(1),n_3(1)]$ results the same $\rho_e$. 
(b) $V_2[\rho_e]$ is the potential whose ground-state configuration results the same $\rho_e$ 
of (a) and is shown along with $V_1[\tilde{\rho}^{(1)}_0]$. (c) $\rho_e[n_1(2),n_3(1),n_4(1)]$ 
is the excited-state density of $1D$ potential well with ground-state $\rho_0$ and produced in 
an alternative configuration $[n_1(2),n_2(1),n_4(1)]~(V_1[\tilde{\rho}^{(1)}_0])$ besides the 
ground-state configuration leading to $V_2[\rho_e]$. (d) Shows all the alternative potentials of 
(c).}
\label{csfig1}
\end{figure}

\begin{figure}[h]
\begin{center}
\includegraphics[width=3.3in,height=3.0in,angle=0.0]{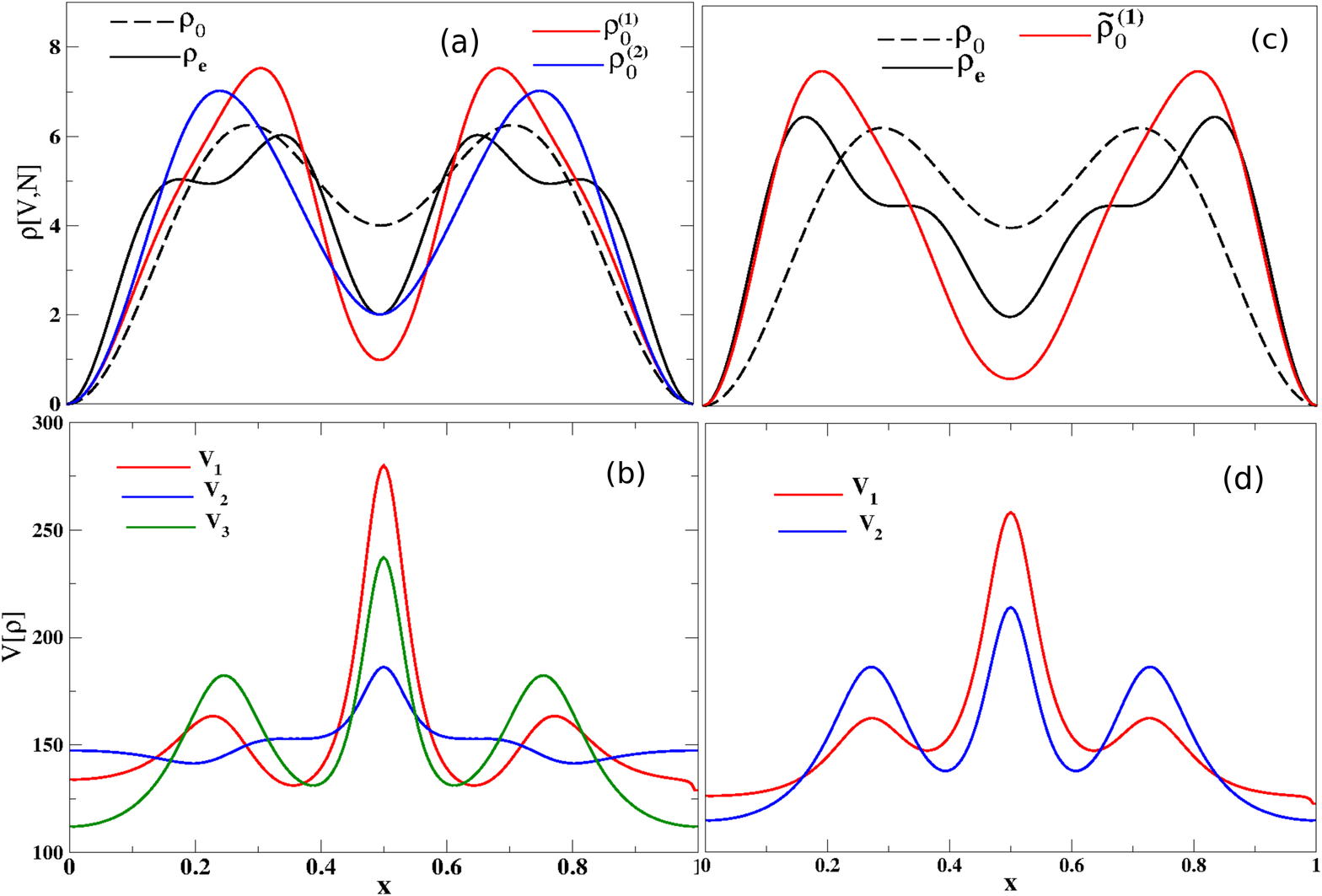} 
\end{center}
\caption{(a) $\rho_e[n_1(1),n_2(2),n_4(1)]$ is the excited-state density of $1D$ potential well 
with ground-state $\rho_0 $. $\tilde{\rho}^{(1)}_0$ and $\tilde{\rho}^{(2)}_0$ are the ground
-state densities of $V_1$ and $V_2$ whose excited-state configurations $[n_1(2),n_2(1),n_3(1)]$ 
and $[n_1(2),n_2(1),n_4(1)]$ results the same $\rho_e$. (b) $V_3[\rho_e]$ is the potential 
whose ground-state configuration gives the same $\rho_e$ of (a) and is shown along with $V_1$, 
$V_2$. (c) $\rho_e[n_2(2),n_3(1),n_4(1)]$ is the excited-state density produced in alternative 
configuration $[n_1(2),n_2(1),n_3(1)]~(V_1[\tilde{\rho}^{(1)}_0])$, besides the ground-state 
configuration leading to $V_2[\rho_e]$. (d) Shows all the alternative potentials in (c).}
\label{csfig2}
\end{figure}

\begin{figure}[h]
\begin{center}
\includegraphics[width=3.3in,height=3.0in,angle=0.0]{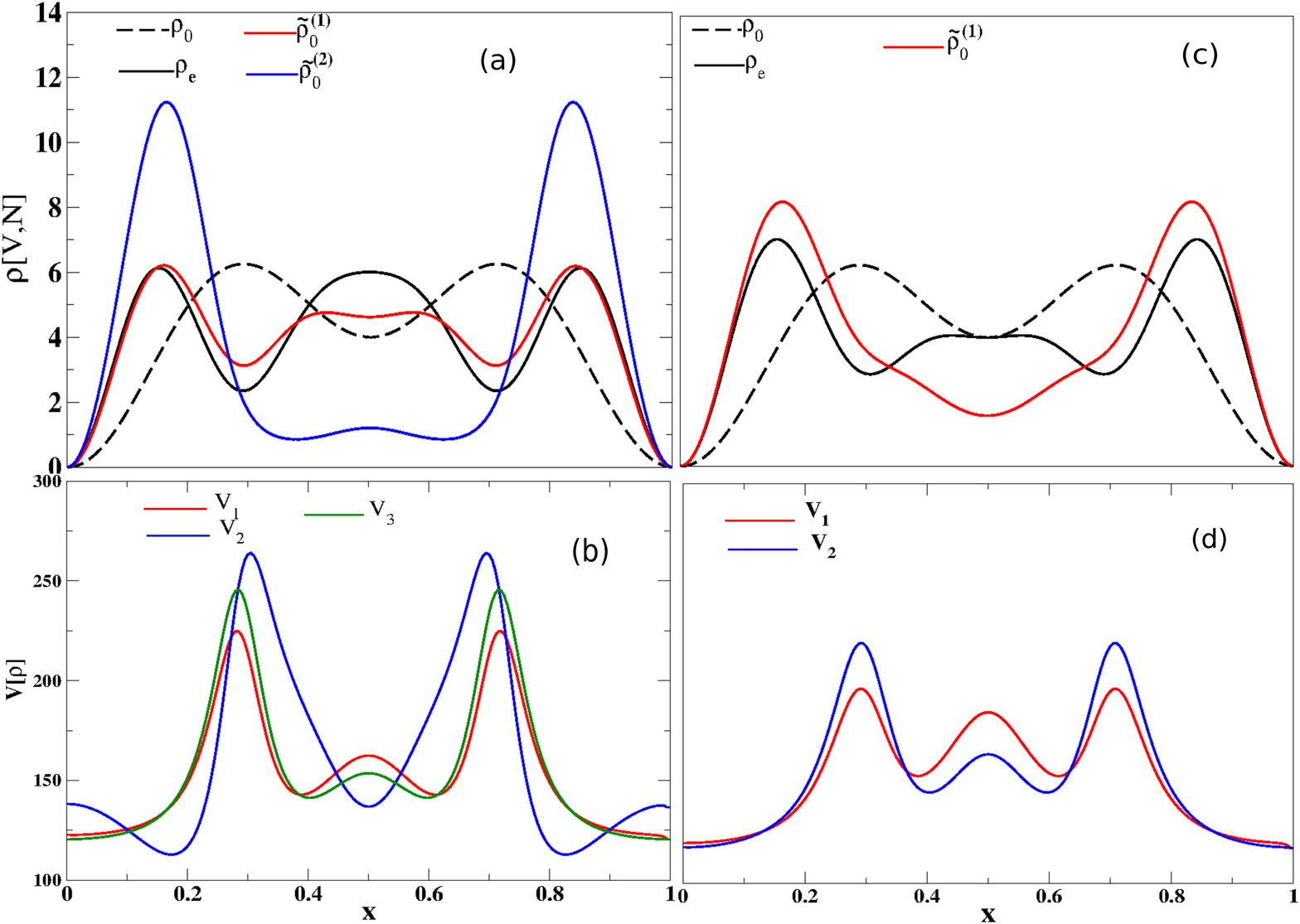} 
\end{center}
\caption{(a) $\rho_e[n_1(1),n_3(2),n_4(1)]$ is the excited-state density of $1D$ infinite potential 
well with ground-state $\rho_0 $. $\tilde{\rho}^{(1)}_0$ and $\tilde{\rho}^{(2)}_0$ are the 
ground-state densities of $V_1$ and $V_2$, whose excited-state configurations $[n_1(2),n_2(1),n_3(1)]$ 
and $[n_1(2),n_3(1),n_4(1)]$ results the same $\rho_e$. (b) $V_3$ is the potential whose ground-state 
density is same as $\rho_e$ of (a) and is shown along with $V_1$, $V_2$. (c) $\rho_e[n_2(1),n_3(2),
n_4(1)]$ is the excited-state density produced via the alternative configurations $[n_1(2),n_2(1),
n_3(1)]~(V_1[\tilde{\rho}^{(1)}_0])$ besides the ground-state configuration leading to $V_2[\rho_e]$. 
(d) Shows all the alternative potentials of (c).}
\label{csfig3}
\end{figure}

\begin{figure}[h]
\begin{center}
\includegraphics[width=3.3in,height=3.0in,angle=0.0]{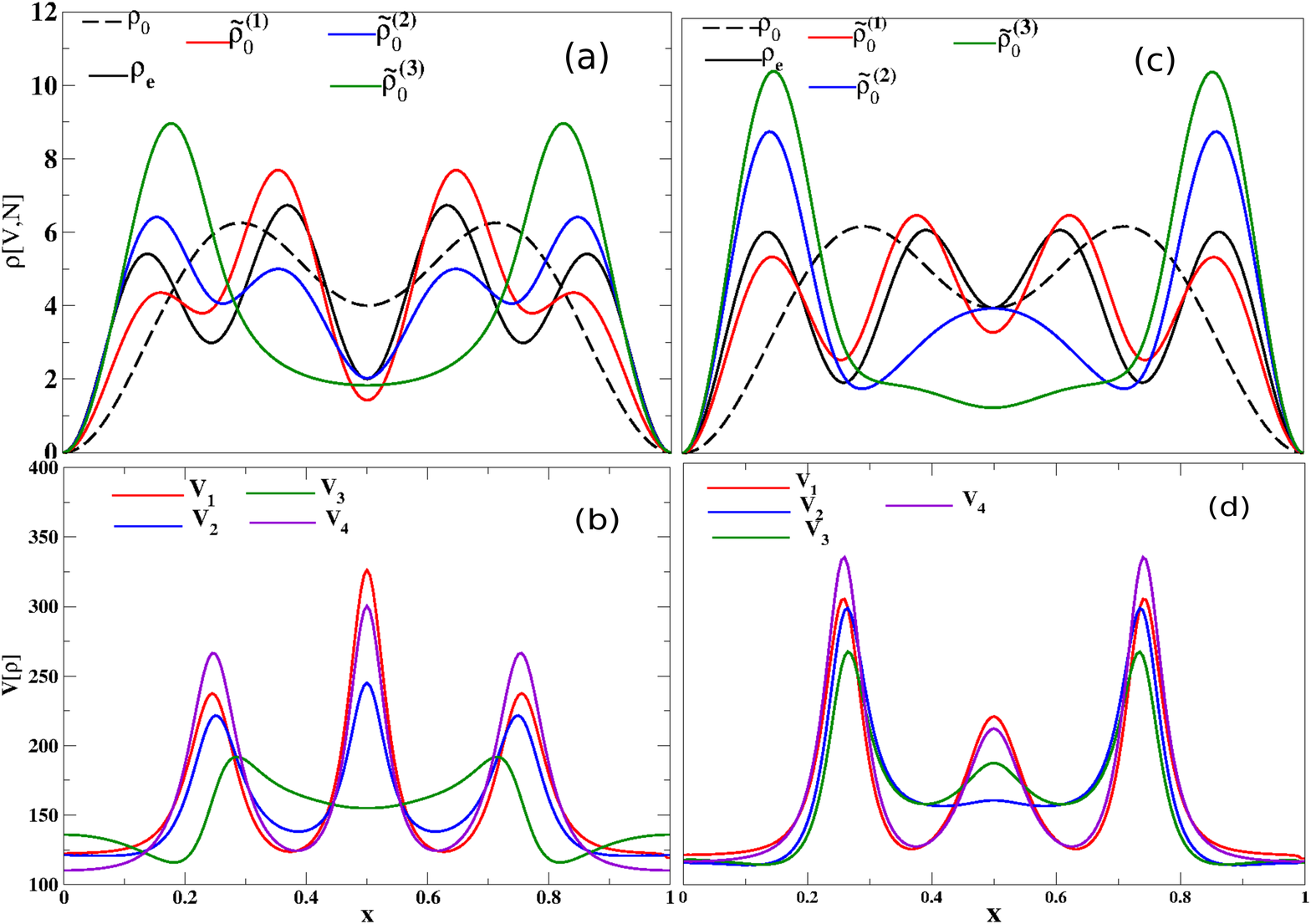} 
\end{center}
\caption{(a) $\rho_e[n_1(1),n_2(1),n_4(2)]$ is the excited-state density of $1D$ infinite potential 
well with ground-state $\rho_0 $. $\tilde{\rho}^{(1)}_0$, $\tilde{\rho}^{(2)}_0$ and $\tilde{\rho}^
{(3)}_0$ are the ground-state densities of $V_1$, $V_2$ and $V_3$ whose excited-state configurations 
$[n_1(2),n_2(1),n_3(1)]$,$[n_1(2),n_2(1),n_4(1)]$ and $[n_1(2),n_4(2)]$ results the same $\rho_e$.  
(b) $V_4$ is the potential whose ground-state density is same as $\rho_e$ of (a) and is shown along 
with $V_1$, $V_2$ and $V_3$. (c) $\rho_e[n_1(1),n_3(1),n_4(2)]$ is the excited-state density 
produced in the alternative configurations $[n_1(2),n_2(1),n_3(1)]~(V_1[\tilde{\rho}^{(1)}_0])$, 
$[n_1(2),n_2(1),n_4(1)]~(V_2[\tilde{\rho}^{(2)}_0])$ and $[n_1(2),n_3(1),n_4(1)]~(V_3[\tilde{\rho}
^{(3)}_0])$ besides the ground-state configuration leading to $V_4[\rho_e]$. (d) Shows all the 
alternative potentials of (c).}
\label{csfig4}
\end{figure}

\begin{figure}[h]
\begin{center}
\includegraphics[width=3.3in,height=3.0in,angle=0.0]{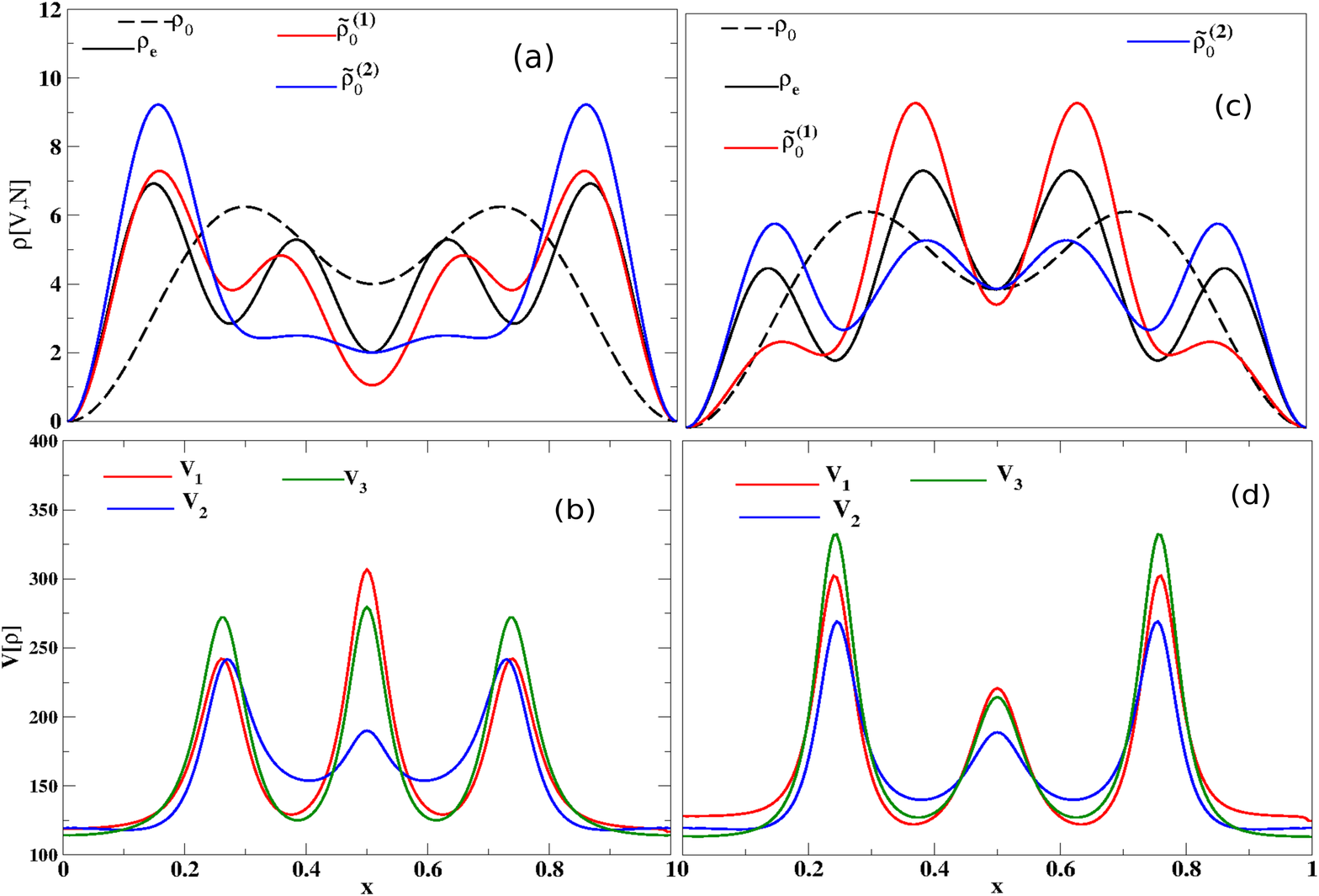} 
\end{center}
\caption{(a) $\rho_e[n_2(1),n_3(1),n_4(2)]$ is the excited-state density of $1D$ infinite potential 
well with ground-state $\rho_0 $. $\tilde{\rho}^{(1)}_0$ and $\tilde{\rho}^{(2)}_0$ are the ground
state densities of $V_1$ and $V_2$, whose excited-state configurations $[n_1(2),n_2(1),n_3(1)]$ 
and $[n_1(2),n_2(1),n_4(1)]$ results the same $\rho_e$. (b) $V_3$ is the potential whose ground
state density is same as $\rho_e$ of (a) and is shown along with $V_1$, $V_2$. (c) $\rho_e[n_1(2),
n_4(2)]$ is the excited-state density produced in alternative configurations $[n_1(2),n_2(1),n_3(1)]
~(V_1[\tilde{\rho}^{(1)}_0])$ and $[n_1(2),n_2(1),n_4(1)]~(V_2[\tilde{\rho}^{(2)}_0])$ besides the 
ground-state configuration leading to $V_3[\rho_e]$. (d) Shows all the alternative potentials of (c).}
\label{csfig5}
\end{figure}

\begin{figure}[h]
\begin{center}
\includegraphics[width=3.3in,height=3.0in,angle=0.0]{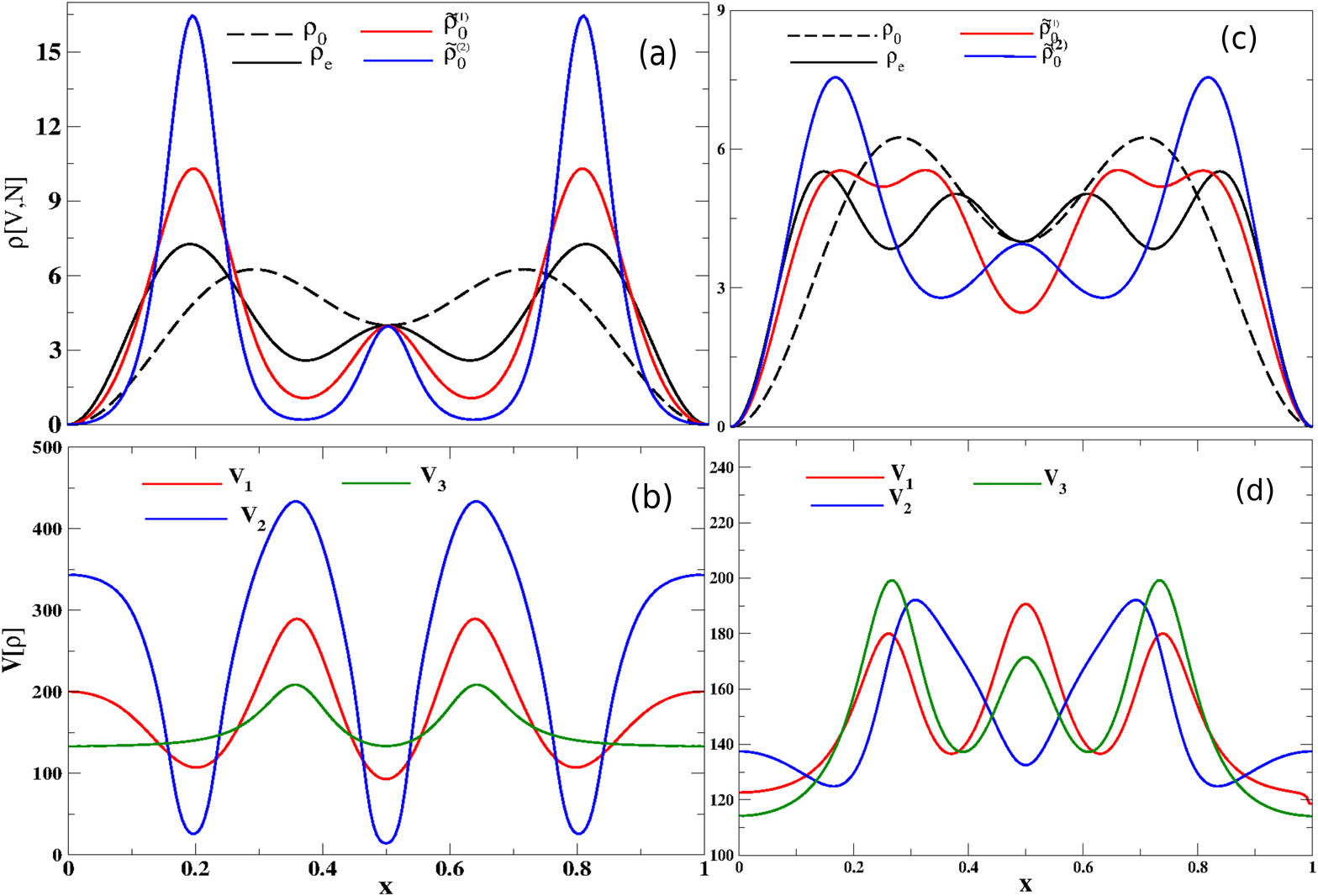} 
\end{center}
\caption{(a) $\rho_e[n_2(2),n_3(2)]$ is the excited-state density of $1D$ infinite potential 
well with ground-state $\rho_0 $. $\tilde{\rho}^{(1)}_0$ and $\tilde{\rho}^{(2)}_0$ are the 
ground-state densities of $V_1$ and $V_2$, whose excited-state configurations $[n_1(2),n_2(1)
,n_4(1)]$ and $[n_1(2),n_4(2)]$ results the same $\rho_e$. (b) $V_3$ is the potential whose 
ground-state density is same as $\rho_e$ of (a) and is shown along with $V_1$, $V_2$. (c) 
$\rho_e[n_1(1),n_2(1),n_3(1),n_4(1)]$ is the excited-state density produced in alternative
configurations $[n_1(2),n_2(1),n_3(1)]~(V_1[\tilde{\rho}^{(1)}_0])$ and $[n_1(2),n_2(1),n_4(1)]
~(V_2[\tilde{\rho}^{(2)}_0])$ besides the ground-state configuration leading to $V_3[\rho_e]$.
(d) Shows all the alternative potentials of (c).} 
\label{csfig6}
\end{figure}

\begin{figure}[h]
\begin{center}
\includegraphics[width=3.0in,height=3.0in,angle=0.0]{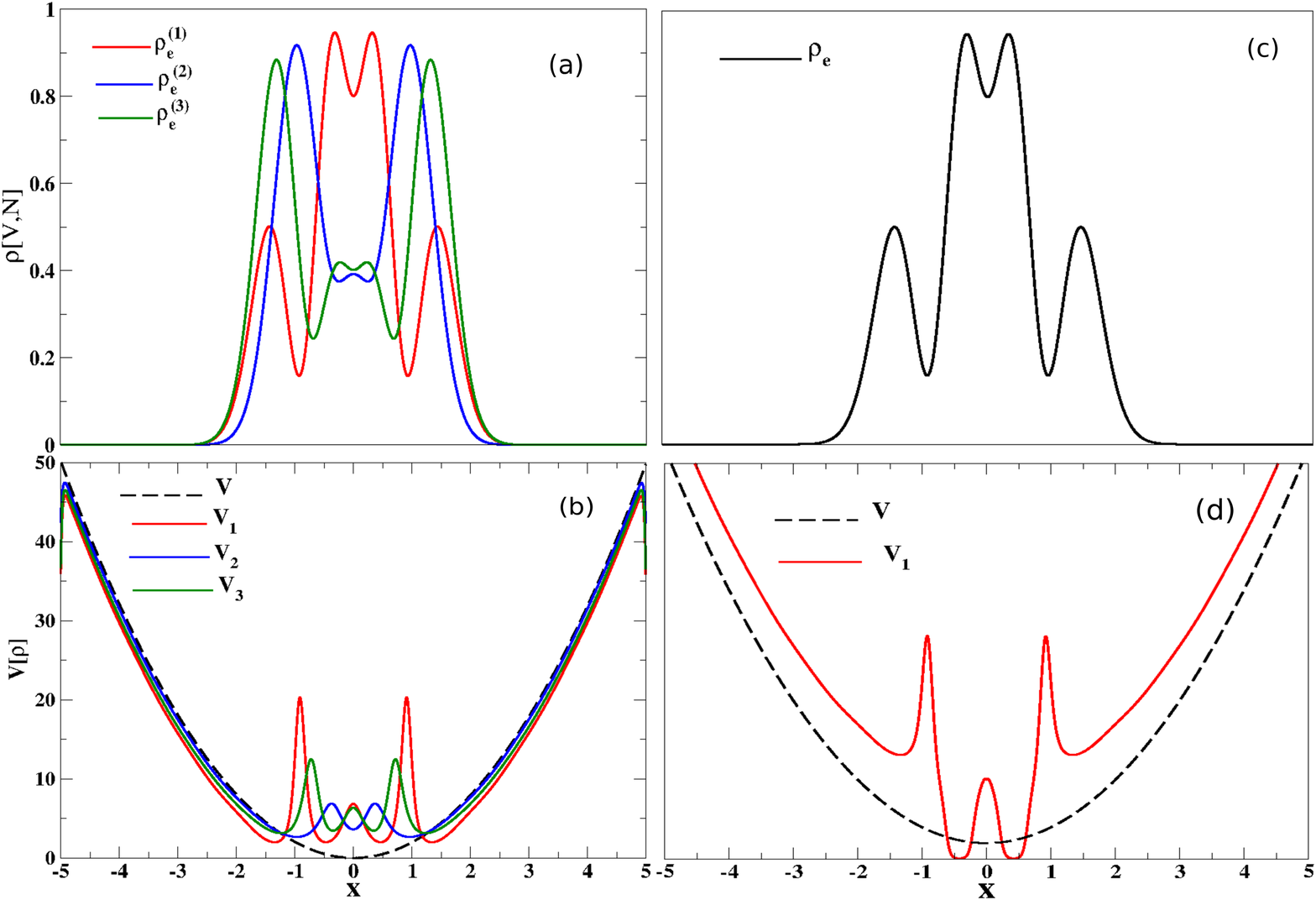} 
\end{center}
\caption{(a) $\rho_e^{(1)} [n=0, n=3]$(both half-filled) , $\rho_e^{(2)} [n=1, n=2]$ (both half 
filled) and $\rho_e^{(3)} [n=2, n=3]$(both half filled) are the excited-state densities of 
the potential $V$ produced as the ground state density of the potentials $V_1$, $V_2$ and $V_3$.
(b) Shows all the four potentials $V$, $V_1$, $V_2$ and $V_3$ of (a). (c) $\rho_e [n=0, n=3]$ 
(both half filled) is the excited-state density of the potential $V$ produced in an alternative 
excited state configuration $[n=0, n=2]~(V_1)$. (d) Shows both the potentials of (c).}
\label{csfig7}
\end{figure}

To conclude this section, we would like to shed some light on the structure of the generated
potentials at the boundary as its very important to be determined accurately. Since the 
wavefunctions die out towards the boundary. Thus the potentials obtained by the Schr\"{o}dinger 
equation inversion (i.e. Eq.\ref{gl5}) for specified eigenvalue differences will attain large 
positive value. Actually, in our approach we have gone way beyond \cite{lpls} to generate 
the accurate structure of the potential which is clear from the results. The important point to 
be noted is that the singularity of the potential plays the most crucial role if one directly 
solving the Schr\"{o}dinger equation. But in getting the potential structure whether by inverting 
the Schr\"{o}dinger equation or CS method solely depends on the wavefunction behavior in a given
domain. So better access of the wavefunction's behavior will by default lead to reliable potential 
structure. 

\section{Results within the CS formalism}
In this section, we will discuss the results in connection with the density-to-potential mapping 
based on the CS-formalism discussed earlier. According to it, there exist multiple potentials 
for a given ground or excited state (eigen)density. But for the case of excited state density, 
when it is produced as some different excited-state of these multiple potentials (except the actual
one) the corresponding ground-states are completely different from that of the original system. 
Similarly, one can produce potentials whose ground-state density may be same as the excited-state 
density of the original system. The results we have obtained for the systems of our study are fully 
consistent with the unified CS $e$DFT. The Zhao-Parr \cite{zp} CS method is being used to show the 
multiplicity of potentials for a given density.

To begin the CS exemplification (shown in Fig.\ref{csfig1}), lets consider $four$ non-interacting 
particles in an $1D$ potential well, where two fermions are in $n=1$ state and one fermion each in 
$n=2$ and $n=4$ state. As a result, this gives some excited state density $\rho_e(x)$ associated 
with the above configuration which is shown in the Fig.\ref{csfig1}(a) and is given by
\begin{equation}
\rho_e(x) = \rho_e^{\text{V}_0}(x) = 2|\Psi_1(x)|^2 + |\Psi_2(x)|^2 + |\Psi_4(x)|^2~,
\label{cseq1}
\end{equation}
where $\Psi_i(x)$~s are the wavefunctions of the $1D$ potential well. In all our results shown in the 
figures (\ref{csfig1}) to (\ref{csfig7}), we have adopted notation $\rho(n_i(f_j))$, where $n_i$ 
denotes the quantum number of the eigenfunctions of the  potential $V$ or $V_i$ ($i = 1, 2, 3, 4$) 
and $f_j$, the occupation. Using CS \cite{zp} the excited state density $\rho_e(x)$ given by Eq.(\ref{cseq1}) 
is produced through another alternative potential $V_1$ (say) whose $n=1$ state is occupied with $2$ 
fermions (i.e. $f_1=2$) and $n=2, n=3$ with one fermion each (i.e. $f_2 = 1 = f_3$). Now the ground 
state density of the potential $V_1$ is different from that of the $V_0$ (i.e. particle in an infinite 
potential well) which is given by $\tilde{\rho}^{(1)}_0$ (Fig.\ref{csfig1}a). As per our formalism, 
there can be many such multiple potentials having the given density as it's eigendensity associated 
with some combination of eigenfunctions. So it is possible that one can also obtain second alternative 
potential $V_2$ (say) whose ground-state density is same as the above excited state density ($\rho_e(x)$) 
of the original system ($V_0$). In this way, we have studied six such excited states of the $1D$ potential 
well (Figs.\ref{csfig1} to \ref{csfig6}) and for each case we are able to produce symmetrically different 
multiple potentials for fix densities. Also in each case, we have produced the alternative potential 
whose ground-state density is nothing but the given excited-state density of the original configuration 
(i.e. $1D$ potential well).

As our final case study, we have considered the excited-states of the $1D$ QHO. This is also an interesting 
model system like the potential well. The results for this case, are shown in Fig.\ref{csfig7}. Now consider 
the Fig.\ref{csfig7}a, in this case we have produced three symmetrically different alternative potentials 
$V_1$, $V_2$ and $V_3$ (shown in Fig.\ref{csfig7}b) whose ground-states densities (i.e. $\rho_0^{(1)}(x), 
\rho_0^{(2)}(x)$ and $\rho_0^{(3)}(x)$)are same as the different excited-states densities (i.e. $\rho_e^{(1)}
(x), \rho_e^{(2)}(x)$ and $\rho_e^{(3)}(x)$) of the QHO potential $V(x)$. Here $\rho_e^{(1)}(x)$ corresponds 
to the configuration [$n=0 (f_0=1), n=3 (f_3=1)$]. Similarly, $\rho_e^{(2)}(x)$ and $\rho_e^{(3)}(x)$ are 
arising from the excited-state configurations [$n=1 (f_1=1), n=2 (f_2=1)$] and [$n=2 (f_2=1), n=3 (f_3=1)$] 
respectively. In Fig.\ref{csfig7}(d), we have produced a different potential $V_1$ whose excited-state density 
corresponding to the configuration [$n=0 (f_0=1), n=2 (f_2=1)$] is same as the excited-state density $\rho_e(x)$ 
([$n=0 (f_0=1), n=3 (f_3=1)$]) of the original $1D$ QHO potential. Although we have produced so many 
potentials, but our criteria will only select the original potentials (i.e. the infinite potential well 
in the previous and QHO in the current study) for any given (i.e. either ground or excited-state) density. 
Thus establishes the excited-state $\rho(x) \Longleftrightarrow {\hat v}(x)$ mapping uniquely.

\section{Discussions}
Now the conceptually basic questions of $e$DFT: what are the consequences as well as similarities and 
differences between the results of the CS formalism and that obtained in connection to the HK/GL 
theorem? Secondly, whether there arisen any critical scenario which is inconsistent with the HK and
/or GL theorem(s)? This is because several multiple potentials are obtained for non-interacting fermions 
in the ground as well as lowest excited state. Not only that, \cite{gb,lpls} have also claimed that for 
higher excited-states there is no analogue of HK theorem. So the seemingly contradictory results may 
give rise to the wrong conclusion about the validity of HK/GL theorem and non-existence of density-to-
potential mapping for excited-states. However, the generalized/unified CS formalism overrules all these
claims by showing that the ground-state density of a given symmetry (potential) can be the excited-state 
density of differing symmetry (potential). Now this excited-state will have a corresponding ground state
which will be obviously quite different from the ground-state of the original system. As a matter of
which there will exist a different potential according to HK theorem. This is also true for the excited
-state density of the actual system: when it becomes either the ground-state or some arbitrary excited
-state density of another potential. So the unified CS formalism justifies the non-violation of HK/GL 
theorem for such states. 

Now based  on the unified/generalized CS $e$DFT, one can very nicely interpret ours as well as \cite{lpls} 
results. Actually by keeping the excited/ground state density fix via a unitary transformation never 
guarantee the symmetries of the states involve will remain intact. This is because by changing the 
$\Delta'$ value and keeping either ground or the excited state density fix, we are forcing the system 
to change itself accordingly without hindering only the fixed density constraint. Since $\Delta'$ is not 
fixed. So in principle one can make several choices for $\Delta'$ and for each choice, the system will
converge to different potentials (systems/configurations) which can give the desired density of ground/
excited-state of the original system (potential/configurations) as one of it's eigendensity. Actually, 
the converged potentials are those for which the G\"{o}rling and LN functionals are stationary and minimum 
respectively. So everything is again automatically fits into realm of generalized CS formalism and nothing 
really contradicting or posing issues for the $e$DFT formulations provided by \cite{gor1,gor2,gor3,ln1,ln2,
ln3,shjp2,thesis,al,aln}. Also the transformed quantum states leading to multitude of potential for a given 
density are energetically far off from the actual system and even the ground-states are also very different. 
Thus the generalized CS formalism proposed in this work along with the SH criteria can be considered as the
most essential steps for establishing the $\rho(\vec r) \Longleftrightarrow {\hat v}_\text{ext}(\vec r)$ 
which further elaborated below. 

Now the question is out of these existing multiple potentials in association with a fix density and 
$\Delta'$, which potential in principle should be picked in view of the $\rho(x) \Longleftrightarrow 
{\hat v}(x)$? The criteria of selecting the exact potential out of all possibilities have already 
been discussed in Sec.III. First of all it is quite obvious from the Figs.(\ref{qhofig0_0-1} $to$ 
\ref{qhofig0_2-2}) and from Fig.\ref{pbfig2-1} $to$ Fig.\ref{pbfig4-3} that the ground-state densities 
of the generated alternate potentials are different from that of the original potential. This is also 
true even for the results of the CS formalism as shown in the Figs.(\ref{csfig1} $to$ \ref{csfig7}). 
So when we are fixing the excited-state density at the same time we should have taken care of the 
ground-state of the newly found system and the old one. Similarly, when several multiple potentials 
are generated for a given ground-state density, the same is not produced as the ground state eigendensity 
of the alternate potentials. So it's quite obvious that there is no violation of the HK theorem. The 
criteria of taking care of the ground-states of the two system is given in Eq.(\ref{cdpm16}). Additionally 
the kinetic energies of the two systems need to be kept closest, which we have pointed out on the basis 
of DVT. So in all the non-interacting model systems reported here, $\Delta T$ should have been zero. But 
the drastically differing structures of the transformed and original wavefunctions are nothing but the 
manifestation of non-vanishing difference of kinetic energies and thus leading to the multiple potentials. 
Furthermore, the most significant differences between the symmetries of the old and new systems implies 
that principally there exist discrepancies in the expectation values of the Hamiltonian w.r.t. the 
ground-states of various multiple potentials. This is what trivially follows from the reported results.
Hence, the proposed criteria uniquely maps a given density of the $1D$ QHO/infinite well to a potential
which is nothing but the $1D$ QHO/infinite well and discards rest of the multiple potentials.

\section{Summary and CONCLUDING REMARKS}
In this work, we have tried to obtain a consistent theory for $e$DFT based on the stationary state, 
variational and GAC formalism of modern DFT. We have provided a unified and general approach for 
dealing with excited-states which follows from previous attempts made by Perdew-Levy, G\"{o}rling,
Levy-Nagy-Ayers and in particular the work of Samal-Harbola in the recent past. In this current attempt,
we have answered the questions raised about the validity of HK and GL theorems to excited-states. 
We have settled the issues by explaining why there exist multiple potentials not only for higher
excited states but also for the ground as well as lowest excited state of given symmetry. In
fact, the existing $e$DFT formalism allows the above possibility and at the same time keeps the 
uniqueness of density-to-potential mapping intact. So we have established in a rigorous fundamental 
footing the non-violation of the HK and GL theorem. Actually, the generalized CS approach gives us a 
strong basis in choosing a potential out of several multiple potentials for a fixed ground/excited
state density. In our propositions, we have strictly defined the bi-density functionals for a fix 
pair of ground and excited-state densities in order to establish the density-to-potential mapping. 
Not only that, the theory also gives us a clear definition of excited-state KS systems through
the comparison of kinetic and exchange-correlation energies w.r.t. the true system. It does takes
care the stationarity and orthogonality of the quantum states. So everything fits quite naturally 
into the realm of modern DFT. 

To conclude, we have demonstrated density-to-potential mapping for non-interacting fermions. For 
interacting case the GAC can be used to formulate all the theoretical and numerical contents in a 
similar way. We are working along this direction for strictly correlated fermions and the results 
will be reported in future. Finally, our conclusion is that nothing really reveals the manifestation 
of the failure or violation of the basic theorems and existing principles of modern DFT irrespective 
of the states under consideration. The method presented by Samal-Harbola and further progress being 
made here provides a most suitable framework and starting ground for the development of new density
-functional methods for the self-consistent treatment of excited states. More realistically, the 
unified CS $e$DFT and further extensions to the SH criteria treat both the ground or excited states 
in an analogous manner. Hence, the present work endows the uniqueness of density-to-potential mapping 
for excited-states with a firm footing.

\section{ACKNOWLEDGMENTS}
The authors thankfully acknowledge valuable discussions with Prof. Manoj K. Harbola and M. 
Hemanadhan.

\end{document}